\documentclass[manuscript]{aastex}

\usepackage{amsbsy,amssymb,amsmath,bm}
\usepackage{graphicx,subfigure}
\usepackage{natbib}
\usepackage{color}

\begin{document}

\title{Magnetic effect on dynamical tide in rapidly rotating astronomical objects}
\author{Xing Wei}
\affil{Institute of Natural Sciences, School of Mathematical Sciences, Shanghai Jiao Tong University}
\email{xing.wei@sjtu.edu.cn}

\begin{abstract}
By numerically solving the equations of rotating magnetohydrodynamics (MHD), the magnetic effect on dynamical tide is studied. It is found that magnetic field has a significant impact not only on the flow structure, i.e. the internal shear layers in rotating flow can be destroyed in the presence of a moderate or stronger magnetic field (in the sense that the Alfv\'en velocity is at least of the order of 0.1 of the surface rotational velocity), but also on the dispersion relation of waves excited by tidal force such that the range of tidal resonance is broadened by magnetic field. A major result is that the total tidal dissipation scales as square of the field strength, which can be used to estimate the strength of internal magnetic field in the astronomical object of a binary system. Moreover, with a moderate or stronger field the ratio of the magnetic dissipation to the viscous dissipation is almost inversely proportional to magnetic Prandtl number (i.e. the ratio of viscosity to magnetic diffusivity), and therefore in the astrophysical situation at small magnetic Prandtl number magnetic dissipation dominates over viscous dissipation with a moderate or stronger field.

\end{abstract}

\maketitle

\section{Introduction}

In a binary system, e.g. binary stars, star and planet, planet and satellite, etc., the angular momentum transfers between the orbital motion and the rotational motion of astronomical objects such that the binary system evolves to the equilibrium state of minimum energy, the so-called synchronization and circularization processes. Tidal dissipation plays an important role in these processes. Tide consists of two parts, equilibrium tide which is a response to the external tidal potential and a slow (compared to the dynamical time scale) quasi-hydrostatic displacement at the surface of astronomical objects \citep{ogilvie-review}, and dynamical tide which is the internal fluid waves resonantly excited by the external tidal force. Equilibrium tide $\xi$ obtained by the hydrostatic balance is equal to the sum of the tidal potential $\Psi$ and the Eulerian perturbation of the self-gravitational potential $\Phi'$ divided by the local gravitational acceleration $g$, i.e. $\xi=-(\Psi+\Phi')/g$ \citep{book-tide,ogilvie-review}. The self-gravitational potential perturbation $\Phi'$ is an order-of-unit multiple of the tidal potential $\Psi$ and the major component of tidal potential is on the ($l=2, m=2$) mode, where $l$ and $m$ are respectively degree and order of spherical harmonics. Therefore, equilibrium tide is on a large length-scale, but the fluid waves of dynamical tide can have small length scales. Energy can dissipate either through the small-scale turbulent convective eddies on the large-scale equilibrium tide \citep{zahn1977,goodman1997} or through the small-scale dynamical tide \citep{zahn1975,goldreich1989}.

There exist various fluid waves in the fluid part (e.g. atmosphere, ocean, interior) of an astronomical object, and these waves have various length and time scales. When the wave frequency is close to the tidal frequency, resonance occurs and the tidal response and dissipation can be greatly improved. At the surface of an astronomical object, gravity induces surface gravity waves, namely f mode (equilibrium tide can be regarded as a forced f mode). The frequency of f mode is of the order of dynamical frequency and f mode is too fast to be resonantly excited by tidal force. In a compressible fluid, pressure induces sound waves, namely p mode, and again p mode is too fast to be resonantly excited by tidal force. In the radiative zone, buoyancy force in a stably stratified layer induces internal gravity waves, namely g mode. The frequency of g mode, i.e. the buoyancy frequency, can be close to tidal frequency and g mode can resonate with external tidal force. Internal gravity waves transport angular momentum and dissipate through radiation. In a rapidly rotating astronomical object, Coriolis force induces inertial waves which can also resonate with tidal force. In the convective zone, inertial waves transport angular momentum and dissipate through small-scale turbulent eddies. Internal gravity wave and inertial wave that resonate with tidal force are both dynamical tide.

Internal gravity wave has been extensively studied. Zahn started this study in his three seminal papers \citeyearpar{zahn1970,zahn1975,zahn1977}, \citet{goldreich1989} gave a physical interpretation and applied to early-type stars (1989), \citet{savonije1983} studied the non-adiabatic effect, \citet{goodman1998} applied to solar-type stars, \citet{lai2012} applied to white dwarfs with a sharp composition jump, etc. Inertial wave has also been extensively studied in \citep{ogilvie2007}, \citep{goodman2009}, \citep{ogilvie2004}, \citep{ogilvie2009}, \citep{ogilvie2013}, \citep{wu2005a}, \citep{wu2005b}, etc. Nonlinear effect on dynamical tide was considered in \citep{goodman1996}, \citep{ogilvie2007}, \citep{ogilvie2010}, \citep{weinberg2012}, \citep{ogilvie2014}, \citep{weinberg2016}, \citep{wei2016a}, etc. The summary and details of the studies about dynamical tide can be found in the recent review paper by \citet{ogilvie-review}.

However, the magnetic effect on dynamical tide is not very clear. \citet{buffett2010} estimated the magnetic (or Ohmic) tidal dissipation in the Earth's fluid core to extrapolate the strength of Earth's magnetic field in the fluid core. He performed numerical calculations for rotating magnetohydrodynamics (MHD) in a spherical shell and assumed that magnetic dissipation focuses in the internal shear layers built by the propagation of inertial waves. In his paper how magnetic field influences the internal shear layers was not considered, i.e. the Lorentz force was neglected in the equation of fluid motion, and the dynamics was completely controlled by rotation. In \citep{wei2016b} I performed a local WKB analysis in rotating MHD about the magnetic effect on dynamical tide, and found two results. The first is that magnetic dissipation wins out viscous dissipation even with a weak magnetic field because the presence of magnetic field modifies the dispersion relation of waves of dynamical tide, i.e. inertial waves with only rotation become magneto-inertial waves with both rotation and magnetic field. The second is that the frequency-averaged dissipation is constant regardless of rotation and magnetic field. But in the local analysis the global spherical geometry was not considered. Most recently, in an online published paper \citep{ogilvie2017} the magnetic effect on dynamical tide was thoroughly studied in the presence of either a uniform vertical or dipolar field in a spherical shell with the vortical forcing method developed by \citet{ogilvie2005}. \citet{ogilvie2017} found the similar results to \citet{wei2016b}, namely magnetic field changes the dissipations because wave dispersion is influenced by magnetic field such that magnetic dissipation can win out viscous dissipation, and the frequency-averaged dissipation is independent of dissipation mechanism. Moreover, they found that a sufficiently strong magnetic field changes flow pattern of internal shear layers, and obtained the scaling law for the competition between rotation and magnetic field, which is very useful for the estimation in astrophysical situation.

In the current paper I will perform numerical calculations for rotating MHD in a spherical shell to study the magnetic effect on dynamical tide, and the impact of magnetic field on flow structure and waves will be considered. Although the problem and the setup in this study are similar to \citep{ogilvie2017}, there are still some differences. The first is that I use the boundary flow method instead of the vortical forcing method and the time-stepping method instead of a super-large matrix solver (matrix is very large especially in the case of strong field). The second is that I study the regime of a stronger field. The third is that I try to find the scaling laws about the tidal dissipation versus the magnetic field strength. In Section 2 the model of this numerical study will be built, in Section 3 the numerical results will be shown and discussed, and in Section 4 a brief summary will be given.

\section{Model}

We use a spherical shell with the radius ratio $R_i/R_o=1/2$ as the model for a spherical astronomical object. The shell rotates rapidly at a constant angular velocity $\bm\Omega=\Omega\hat{\bm z}$, where $\hat{\bm z}$ is the unit vector in the rotational axis, and a conducting fluid resides within the shell to mimic the convective zone of a star or giant planet or the fluid core of a terrestrial planet or satellite. The inner sphere rotates uniformly to mimic the radiative zone of a star or giant planet or the solid inner core of a terrestrial planet or satellite. A uniform vertical magnetic field $\bm B=B\hat{\bm z}$ is imposed throughout the shell and the motion of the conducting fluid in the shell will induce a new field $\bm b$, which is weak compared to $\bm B$. We numerically solve the equations of rotating MHD in a frame rotating at the angular frequency $\Omega$. Because sound waves in a compressible fluid are too fast to be resonantly excited by tidal force, we consider an incompressible fluid to filter out sound waves. This will bring us a big numerical advantage to avoid the very small time-step for capturing sound waves. We study the linear regime and neglect the nonlinear quadratic terms. We solve the dimensionless rather than dimensional equations, because dimensionless equations can give us more, e.g. scaling laws to extrapolate to the real parameter regime which is not accessible under the current computational power.

The dimensionless equation of fluid motion in the rotating frame reads
\begin{equation}\label{eq:ns}
\frac{\partial\bm u}{\partial t}=-\bm\nabla p'+E\nabla^2\bm u+2\bm u\times\hat{\bm z}+S^2(\bm\nabla\times\bm b)\times\hat{\bm z}.
\end{equation}
In Equation \eqref{eq:ns}, the term on the left-hand-side is the acceleration of fluid motion, the terms on the right-hand-side are successively the Eulerian perturbation of pressure gradient incorporating the centrifugal potential perturbation, the viscous force, the Coriolis force arising from rotation and the Eulerian perturbation of Lorentz force arising from magnetic field, and the nonlinear quadratic terms $\bm u\cdot\bm\nabla\bm u$ and $(\bm\nabla\times\bm b)\times\bm b$ are neglected. The tidal force will be implemented by the outer boundary condition. In Equation \eqref{eq:ns}, length is normalized with the outer radius $R_o$, time with the inverse of rotational speed $\Omega^{-1}$, velocity with $\Omega R_o$, and magnetic field with the strength of the imposed field $B$. The two dimensionless numbers are the Ekman number 
\begin{equation}\label{eq:E}
E=\frac{\nu}{\Omega R_o^2}
\end{equation}
which is the ratio of the rotational time-scale $\Omega^{-1}$ to the viscous time-scale $R_o^2/\nu$ ($\nu$ being viscosity) and measures the strength of rotation $\Omega$ relative to viscosity $\nu$, and the $S$ number
\begin{equation}\label{eq:S}
S=\frac{B}{\sqrt{\rho\mu}\Omega R_o}
\end{equation}
which is the ratio of the Alfv\'en speed $B/\sqrt{\rho\mu}$ ($\mu$ being magnetic permeability) to the angular speed $\Omega R_o$ and measures the strength of the imposed field $B$ relative to rotation $\Omega$. The $S$ number is called the Lenhert number \citep{lehnert}.

On the inner boundary, the horizontal velocity is taken to be stress-free to minimize the effect of boundary layer and the radial velocity vanishes, i.e. fluid cannot penetrate across the inner boundary. On the outer boundary, the horizontal velocity is also taken to be stress-free but the radial velocity follows an oscillatory rising-falling equilibrium tide, i.e.
\begin{equation}\label{eq:bc}
u_r=AP_l^m(\cos\theta)\cos(m\phi-\omega t).
\end{equation}
This boundary condition enforces equilibrium tide at the outer surface to excite dynamical tide in the interior. The interior flow driven by the boundary radial flow consists of both tidal flow (equilibrium tide) and waves (dynamical tide). In Equation \eqref{eq:bc}, $A$ is the amplitude and in the linear problem it is taken to be 1, $P_l^m$ is the associate Legendre polynomial normalized by $\sqrt{\frac{(2l+1)(l-m)!}{4\pi(l+m)!}}$ and $l$ and $m$ are taken to be those of the major component of tidal potential, namely ($l=2, m=2$), $\theta$ and $\phi$ are respectively colatitude and longitude in the spherical polar coordinates, and $\omega$ is the tidal frequency in the rotating frame, i.e. the Doppler shifted frequency $\omega=\omega_i-m\Omega$ ($\omega_i$ being the tidal frequency in the inertial frame). The method to use equilibrium tide to excite dynamical tide was proposed and validated by \citet{ogilvie2009} and has been numerically implemented in \citep{ogilvie2014}. 

\citet{campbell} studied the magnetic effect on stellar oscillations and found that a strong magnetic field induces a boundary layer to modify the radial flow at the surface. For stellar oscillations (free-oscillation problem) magnetic field is significant for boundary condition whereas for tide (forced-oscillation problem) it is not so significant because the tidal frequency is too slow compared to the dynamical frequency at the surface, i.e. the low-frequency limit. We now illustrate this point. The normal stress at the disturbed free surface vanishes, i.e. 
\begin{equation}
-(p'-\rho g\xi_r)-\frac{2\bm B\cdot\bm b}{2\mu}+2\rho\nu\frac{\partial u_r}{\partial r}+\frac{2B_rb_r}{\mu}=0.
\end{equation}
The terms are successively the Eulerian perturbation of fluid pressure, the excess pressure due to the disturbed surface, the Eulerian perturbation of magnetic pressure, the viscous normal stress, and the Eulerian perturbation of magnetic normal stress. The round bracket consisting of the first two terms is the Lagrangian perturbation of pressure. In the Eulerian perturbations of the magnetic pressure and normal stress, the quadratic terms $b^2/2\mu$ and $b_r^2/\mu$ are neglected in the linear regime. In \citep{ogilvie2009} it has been verified that in the low-frequency limit $\omega\ll\omega_d\sim(g/R)^{1/2}$, the Eulerian perturbation of fluid pressure and the viscous normal stress are negligible such that the radial displacement is the equilibrium tide with a correction due to self-gravitation. This result is in agreement with \citep{goldreich1989}. Thus the radial flow at the free surface is proportional to the tidal potential. We now show that the Eulerian perturbation of magnetic pressure is negligible in the low-frequency limit. The ratio of the Eulerian perturbation of magnetic pressure to the excess pressure is
\begin{equation}
\left|\frac{2\bm B\cdot\bm b}{2\mu}\right|\frac{1}{|\rho g\xi_r|}\sim\frac{B^2}{\rho\mu(\Omega R)^2}\frac{b}{B}\frac{(\Omega R)^2}{(R\omega_d^2)\xi_r}=S^2\frac{b}{B}\left(\frac{\Omega}{\omega_d}\right)^2\frac{R}{\xi_r}.
\end{equation}
The radial flow $u_r$ at surface is $u_r=\partial\xi_r/\partial t\sim\omega\xi_r$, and on the other hand $u_r\sim A\Omega R$ where $A$ is the dimensionless factor for the boundary radial flow \eqref{eq:bc}. We are then led to
\begin{equation}
\left|\frac{2\bm B\cdot\bm b}{2\mu}\right|\frac{1}{|\rho g\xi_r|}\sim S^2\frac{b}{B}\frac{\Omega}{\omega_d}\frac{\omega}{\omega_d}\frac{1}{A}<S^2\frac{\Omega}{\omega_d}\frac{\omega}{\omega_d}\frac{1}{A}.
\end{equation}
In the usual astrophysical situations, magnetic field is not so strong that $S$ is less than unity (in our calculations $S^2$ is up to 0.2), rotation frequency $\Omega$ is much less than dynamical frequency $\omega_d$, tidal frequency $\omega$ is also much less than dynamical frequency $\omega_d$, i.e. the low-frequency limit. Although the factor $A$ which measures the strength of equilibrium tide is very small, it is still much larger than $S^2(\Omega/\omega_d)(\omega/\omega_d)$. Therefore this ratio is very small and the magnetic effect at surface is negligible. Similarly, the Eulerian perturbation of magnetic normal stress is also much less than the excess pressure. Consequently, in the tidal problem magnetic field is negligible at surface because of the low-frequency limit. The radial-flow boundary condition is valid to study the magnetic effect on tide. However, it should be noted that in some situations this ratio is not so small. For example, a star or planet in binary system rotates so fast that $\Omega/\omega_d\sim 0.1$, the orbital motion is so fast that $\omega/\omega_d\sim 0.1$ (breaking the low-frequency limit), and the surface displacement is so small that $A\sim 0.01$. If magnetic field at surface is so strong or density at surface is so low that $S^2$ is of order of unity, then this ratio will be of order of unity at surface. Therefore, in the situation of fast rotation, fast orbital motion, strong surface field, and low surface density, magnetic field will have a significant effect at surface and it even deforms the spherical geometry. In our calculations we assume this ratio is much less than unity in the low-frequency limit and neglect the magnetic effect at surface.

The dimensionless equation of magnetic field reads
\begin{equation}\label{eq:b}
\frac{\partial\bm b}{\partial t}=\bm\nabla\times(\bm u\times\hat{\bm z})+\frac{E}{Pm}\nabla^2\bm b,
\end{equation}
where the nonlinear term $\bm\nabla\times(\bm u\times\bm b)$ is neglected and the magnetic Prandtl number
\begin{equation}\label{eq:pm}
Pm=\frac{\nu}{\eta}
\end{equation}
is the ratio of viscosity $\nu$ to magnetic diffusivity $\eta$. For the magnetic boundary condition, the magnetic field in the insulating exteriors of both the $r>R_o$ and $r<R_i$ regions is required to match a potential field.

The numerical method is the standard spectral method with spherical harmonics used on the spherical surface and Chebyshev polynomials used in the radial direction, and the toroidal-poloidal decomposition method is employed for divergence-free condition of velocity and magnetic field \citep{hollerbach2000}. The time-stepping calculation is used to find the final `steady' state (here `steady' is in terms of the volume integrals of energy and dissipation). An alternative method is to separate $\omega$ (namely $\partial/\partial t=i\omega$) and then to solve a super-large matrix problem, and we do not use it because it is numerically difficult in the MHD case. In the longitude direction only one mode $m=2$ is necessary for the linear problem, but in the latitude direction the modes much higher than $l=2$ should be involved because the Coriolis force, the Lorentz force and the induction term will couple more modes in this direction. In our numerical calculations the sufficient resolution for convergence is guaranteed that both the radial and latitude spectra of both kinetic and magnetic energies span more than 10 magnitudes.

In our calculations we focus on the magnetic effect on dynamical tide so that we investigate the two parameters related to magnetic field, $S$ and $Pm$, as well as the tidal frequency $\omega$ but keep all the other parameters constants. The radius ratio $R_i/R_o$ is taken to be $1/2$, which is the value chosen in \citep{ogilvie2014}. A larger radius ratio leads to the stronger excitation of inertial waves. We choose this value for comparison with the results in \citep{ogilvie2014} to validate our numerical calculations. The Ekman number is taken to be $E=10^{-4}$ which is already sufficiently low for the study of rapid rotation (too low Ekman number is numerically demanding). $l$ and $m$ are taken to be $2$. The frequency of inertial waves is lower than $2\Omega$ such that the tidal frequency higher than $2\Omega$ cannot excite inertial waves in rotating flow. However, in the presence of magnetic field the frequency of magneto-inertial waves can exceed $2\Omega$ \citep{wei2016b}, and so the regime of tidal frequency higher than $2\Omega$ will be studied. In our calculations the dimensionless tidal frequency $\omega$ ranges from 0.1 to 3 with the spacing 0.1, i.e. 30 values for $\omega$ in our calculations. We will calculate 6 values for $S$, i.e. $S^2=0$, $0.01$, $0.02$, $0.05$, $0.1$ and $0.2$. $S=0$ is the case of rotating flow without magnetic field. At $S=\sqrt{0.2}\approx 0.45$ magnetic field is already sufficiently strong to be comparable to rotation (see Equation \eqref{eq:S}). We will calculate 4 values for $Pm$, i.e. $1$, $0.5$, $0.2$ and $0.1$, i.e. spanning one magnitude. In the astrophysical situation $Pm$ is very low, e.g. $Pm$ is of the order of $10^{-6}$ in the Earth's liquid-iron core or in the Jupiter's conducting region or in the Sun's convection zone. Too low $Pm$ is numerically demanding and so we study the range of one magnitude to try to find scaling laws. Altogether we have $30\times(6+4)=300$ calculations that will be presented in the next section.

\section{Results}

By numerically solving Equations \eqref{eq:ns} and \eqref{eq:b} with the boundary condition \eqref{eq:bc}, we obtain velocity $\bm u$ and induced field $\bm b$. Kinetic energy indicates the tidal response and the kinetic and magnetic dissipations are responsible for the orbital evolution of a binary system. Then we calculate the volume integrals of kinetic energy and the two dissipations over the spherical shell. In the dimensionless expressions, kinetic energy is normalized with $\rho R_o^5\Omega^2$ and both dissipations with $\rho\nu R_o^3\Omega^2$, and thus
\begin{align}\label{eq:energy}
\mbox{kinetic energy}&=\frac{1}{2}\int|\bm u|^2dV, \nonumber\\
\mbox{viscous dissipation}&=2\int S_{ij}S_{ij}dV, \\
\mbox{magnetic dissipation}&=\frac{S^2}{Pm}\int|\bm\nabla\times\bm b|^2dV, \nonumber
\end{align}
where $S_{ij}$ is the strain tensor. In the next of this section we will use the dimensionless expressions \eqref{eq:energy} as output. 

We also calculate the total angular momentum and it conserves in our linear model. It should be noted that in \citep{ogilvie2014} angular momentum conserves in the linear model of rotating flow but does not conserve in the nonlinear model, and in our linear model of rotating MHD it conserves as well. It may be inferred that angular momentum does not conserve in the nonlinear model of rotating MHD, which we do not study in this paper. We need to remind readers that this method to use equilibrium tide at the outer surface to excite dynamical tide in the interior is applicable in the linear regime but not in the nonlinear regime.

\subsection{Investigation of $S$}

\begin{figure}
\centering
\includegraphics[scale=0.45]{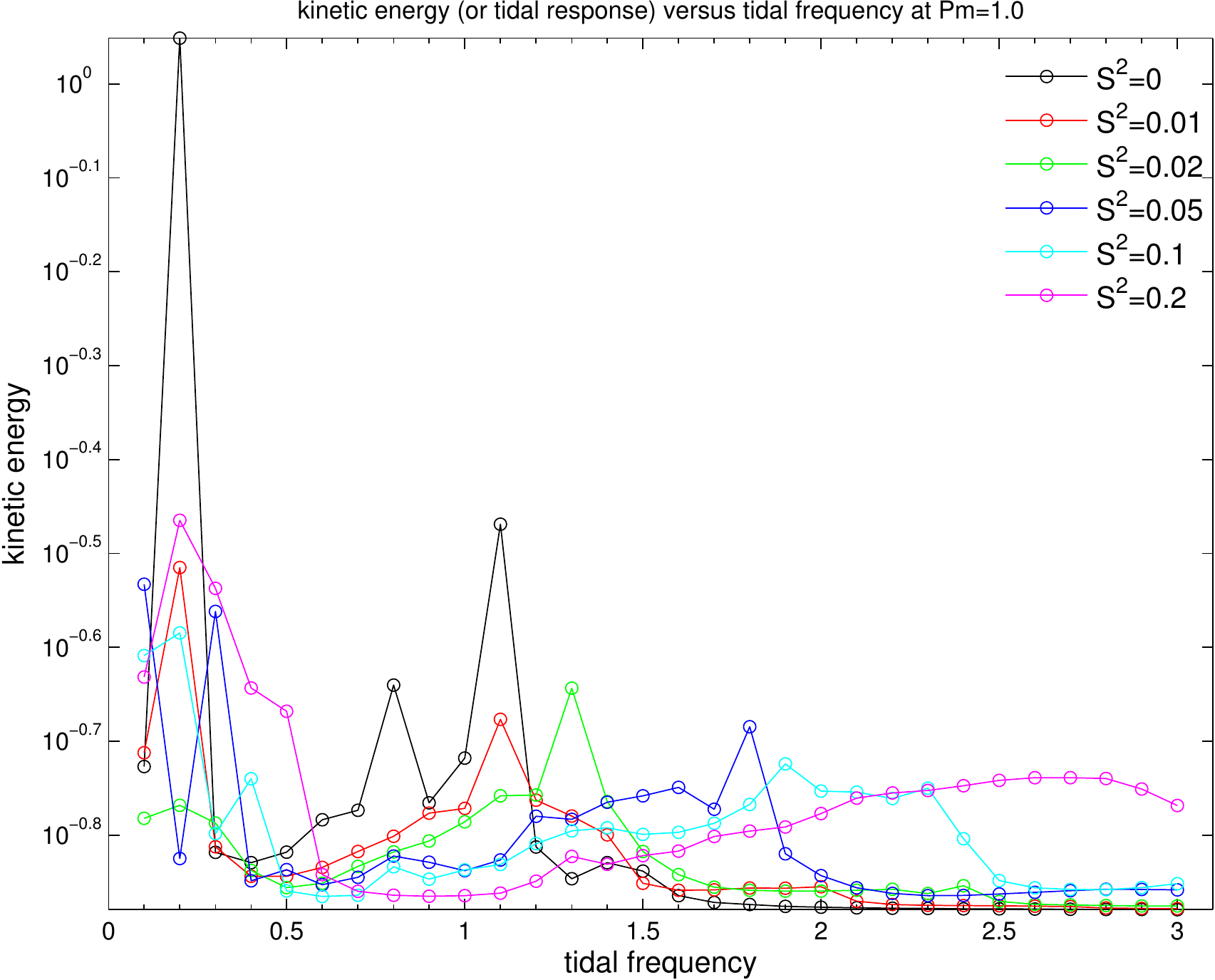}
\caption{Kinetic energy (or tidal response) versus tidal frequency at various $S^2$. $Pm=1.0$.}\label{fig1}
\end{figure}

\begin{figure}
\centering
\subfigure[]{\includegraphics[scale=0.45]{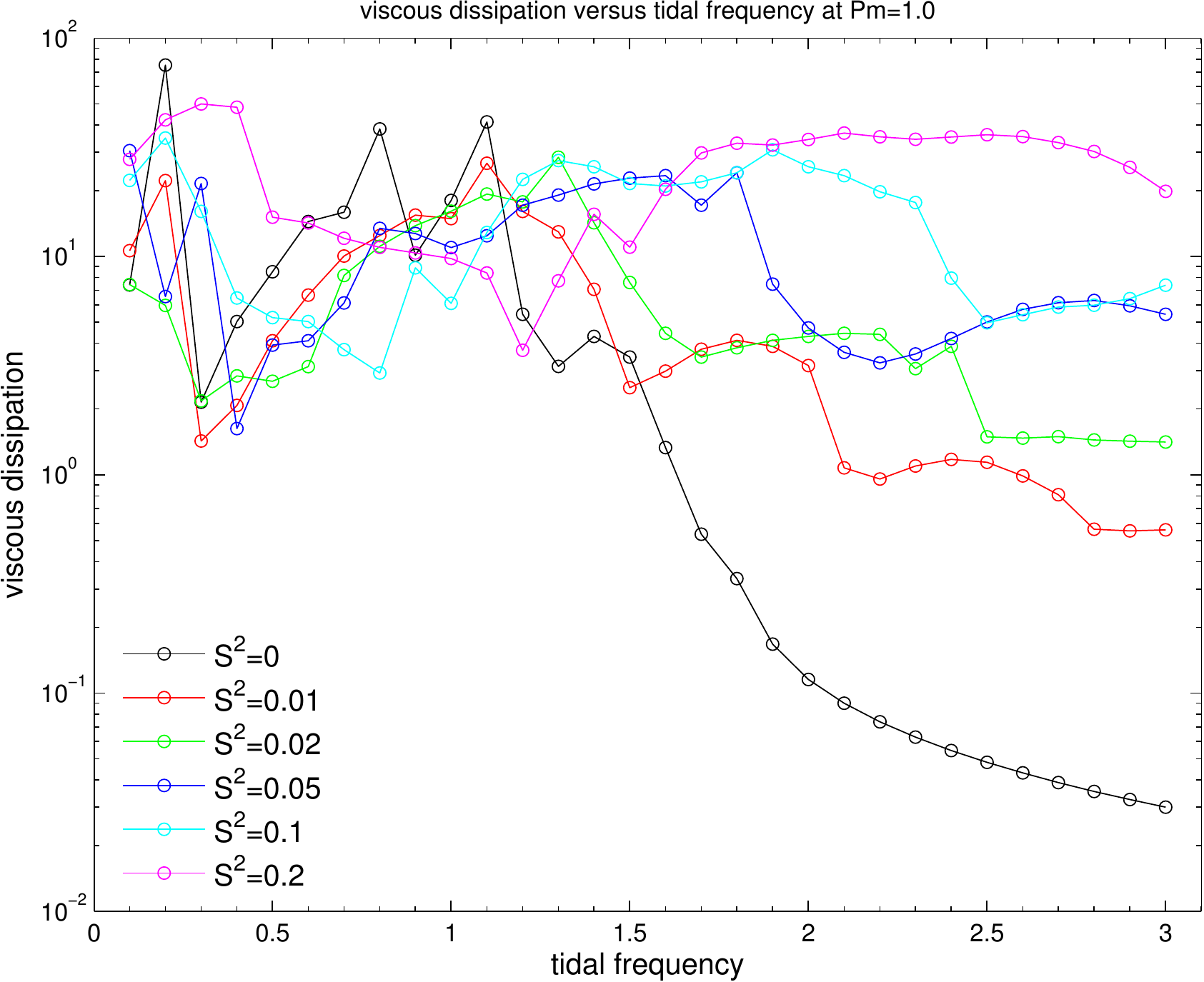}\label{fig2a}}
\subfigure[]{\includegraphics[scale=0.45]{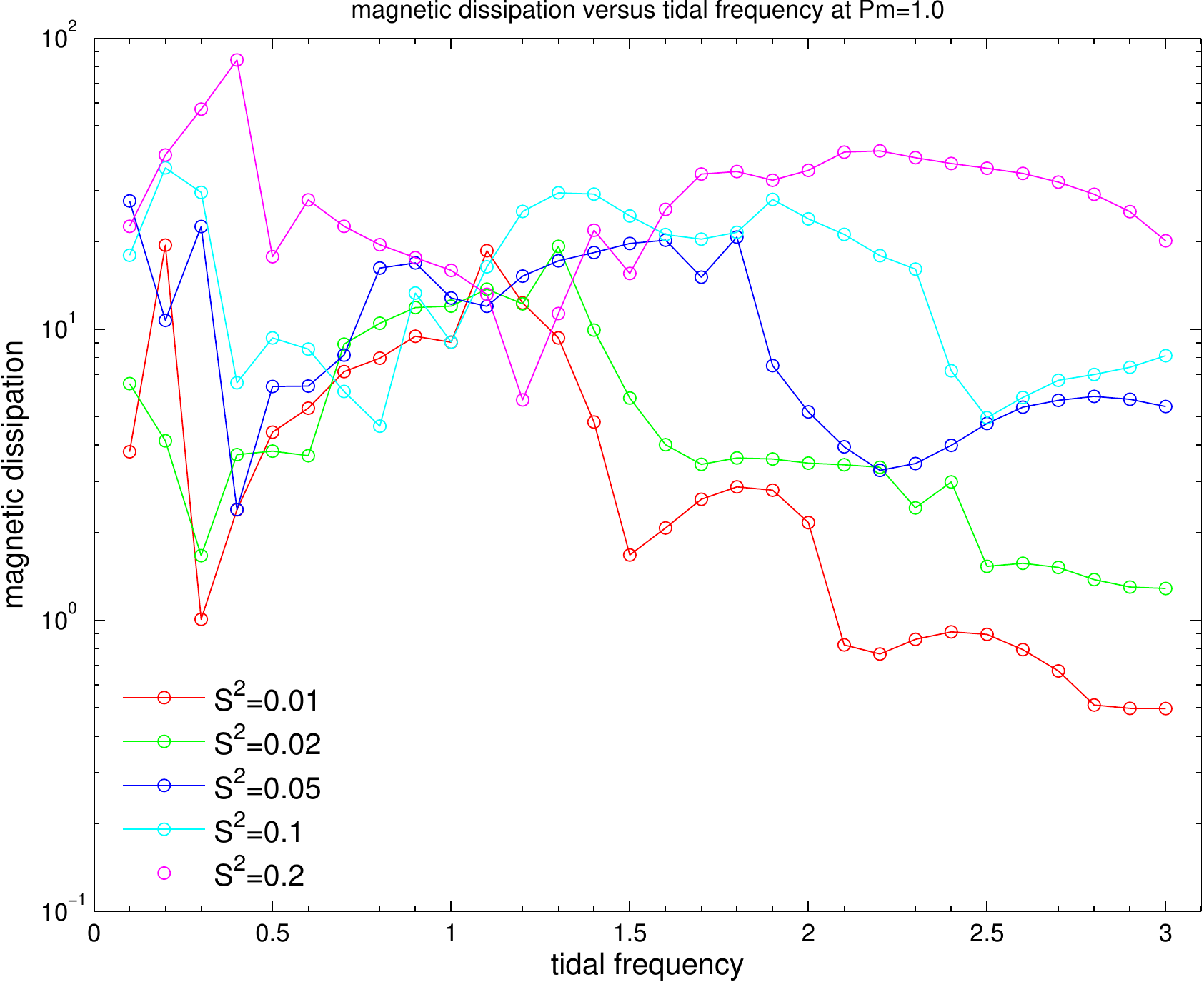}\label{fig2b}}
\subfigure[]{\includegraphics[scale=0.45]{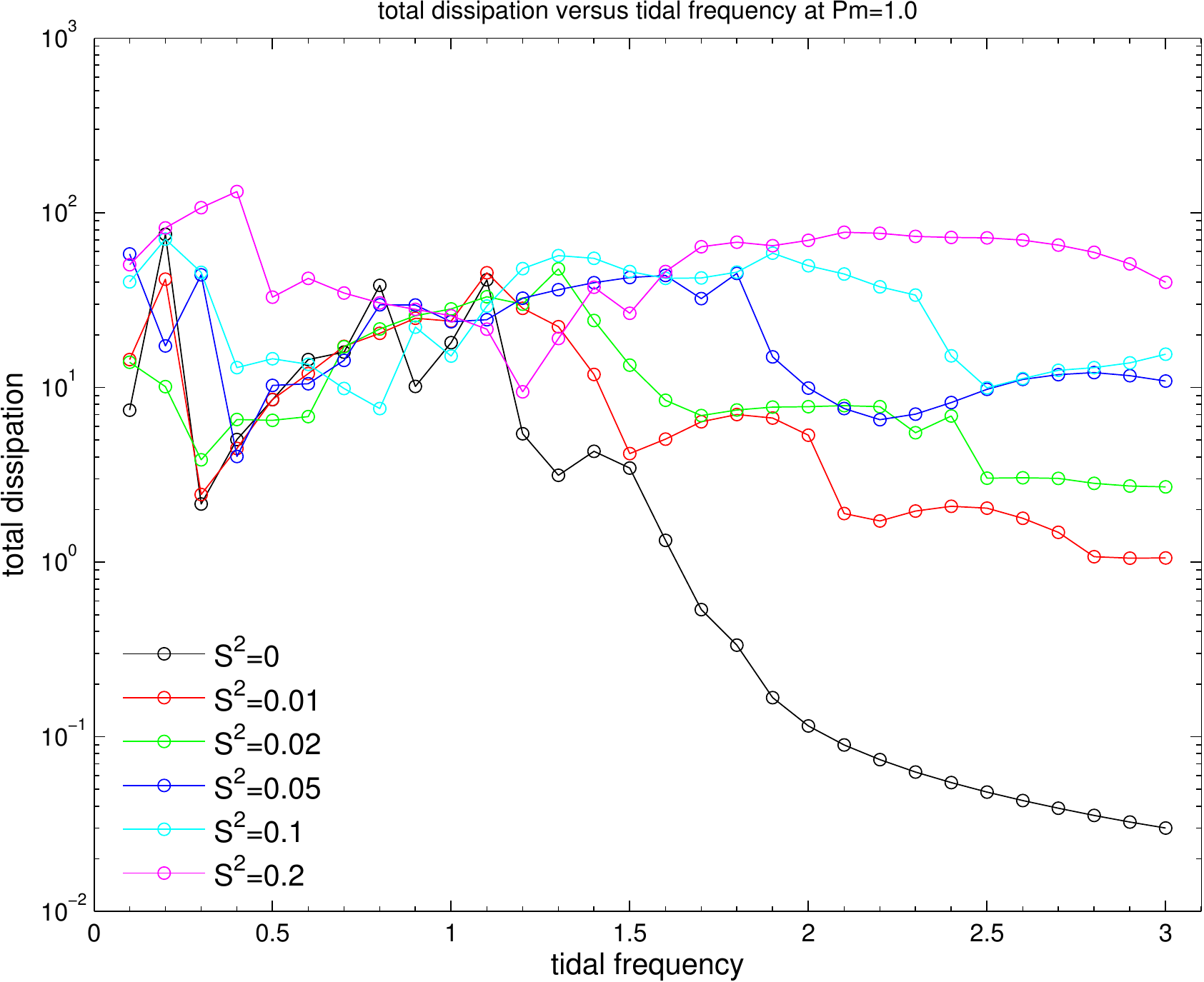}\label{fig2c}}
\subfigure[]{\includegraphics[scale=0.45]{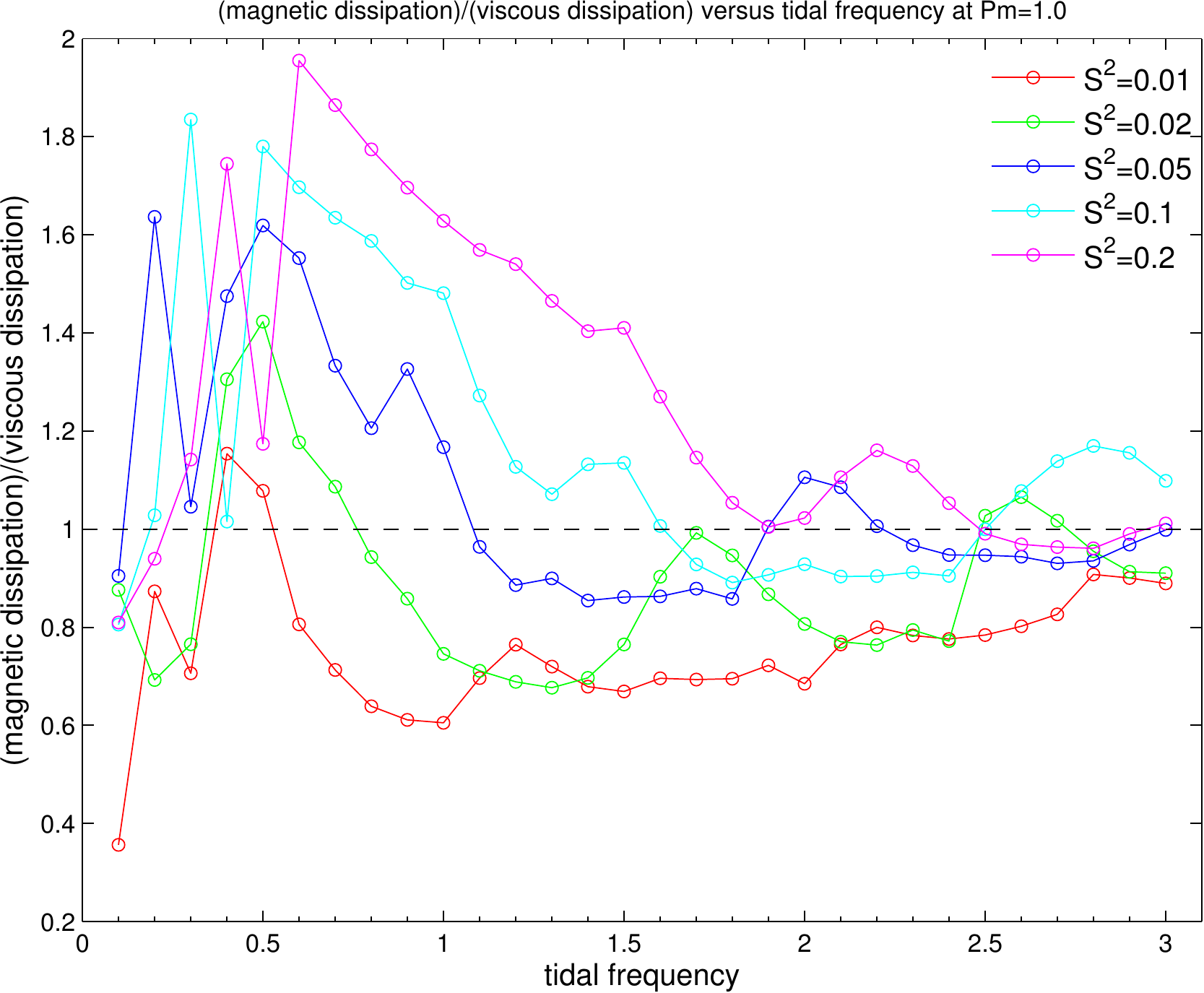}\label{fig2d}}
\caption{(a) Viscous dissipation, (b) magnetic dissipation, (c) total dissipation and (d) ratio of magnetic to viscous dissipations versus tidal frequency at various $S^2$. $Pm=1.0$.}\label{fig2}
\end{figure}

Firstly, we investigate the strength of imposed magnetic field, namely the dimensionless number $S$. Figure \ref{fig1} shows the kinetic energy as the tidal response versus frequency at various $S^2$ with $Pm$ fixed to be 1. The black curve at $S=0$ denotes the kinetic energy in rotating flow. The peak values at the tidal frequencies $\omega=0.2$, $0.8$, $1.1$, $1.4$ are the ones near the resonance of inertial waves and tidal force. It should be noted that the tidal frequencies at these peaks are not exactly but near the eigen-frequencies of inertial waves. To obtain the accurate eigen-frequenices, as illustrated in Section 2, a super-large matrix problem should be solved, which is numerically difficult in the MHD case. It is well-known that the frequency of inertial waves is lower than $2\Omega$, and therefore, on the black curve at the frequencies higher than 2 there does not exist any tidal response of inertial waves. When magnetic field is present and its strength ($S$) increases, some new peaks appear, e.g. $\omega=1.3$ at $S^2=0.02$, $\omega=0.3$ and $1.8$ at $S^2=0.05$, $\omega=0.4$ and $1.9$ at $S^2=0.1$, etc. This is because magnetic field modifies the dispersion relation of waves, i.e. inertial waves in rotating flow become magneto-inertial waves in rotating MHD, such that the eigen-frequencies are changed and the tidal resonances appear at different frequencies. Moreover, with a strong magnetic field, e.g. at $S^2=0.1$ and $0.2$, there exist the tidal responses at the tidal frequencies higher than 2. These results indicate that magnetic field broadens the range of tidal resonance in the global spherical geometry.

Figure \ref{fig2a} shows the viscous dissipation. For the tidal frequency higher than 2 ($\omega>2$), the viscous dissipation is very small in rotating flow ($S=0$) because no inertial wave can be excited in this range of tidal frequency, but it is moderate and increases with the field strength increasing in this range. In the presence of a strong field, e.g. $S^2=0.2$, the viscous dissipation for $\omega>2$ can be even higher than that for $\omega<2$. Figure \ref{fig2b} shows the magnetic dissipation. Its dependence on the tidal frequency is similar to that of viscous dissipation. Figure \ref{fig2c} shows the total dissipation, namely viscous dissipation + magnetic dissipation. Not surprisingly, it is very small for $\omega>2$ in rotating flow but does not vary significantly with tidal frequency in the presence of a strong field. Figure \ref{fig2d} shows the ratio of magnetic dissipation to viscous dissipation. The dashed line indicates the equipartition of the two dissipations. When the tidal frequency increases the two dissipations tend to be close to each other and in the high-frequency range (approximately $\omega>2$) the two dissipations are almost equal to each other. But in the low-frequency range (approximately $\omega<2$) magnetic dissipation wins out viscous dissipation with a strong field and viscous dissipation wins out magnetic dissipation with a weak field.

The result of Figures \ref{fig1} and \ref{fig2} can be applied to the astrophysical situation. {\bf Magnetic field broadens the range of tidal resonance} such that the tidal force with its frequency higher than $2\Omega$ which cannot lead to significant viscous dissipation without magnetic field can lead to significant viscous and magnetic dissipations with magnetic field. For example, in a binary system, if the orbital frequency in the rotating frame is faster than twice of the rotational frequency then inertial waves cannot be excited by the tidal force, but in the presence of a strong magnetic field the tidal force can excite magneto-inertial waves and energy can dissipate very quickly through both viscous and magnetic dissipations.

\begin{figure}
\centering
\subfigure[]{\includegraphics[scale=0.4]{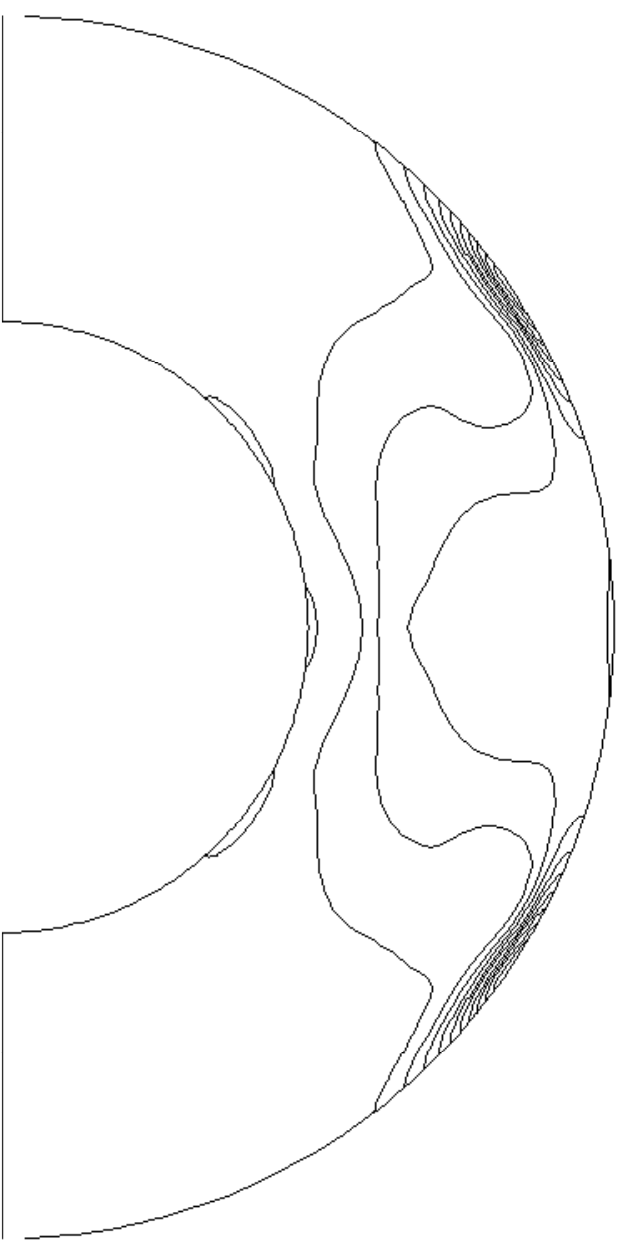}\label{fig3a}} \\
\subfigure[]{\includegraphics[scale=0.4]{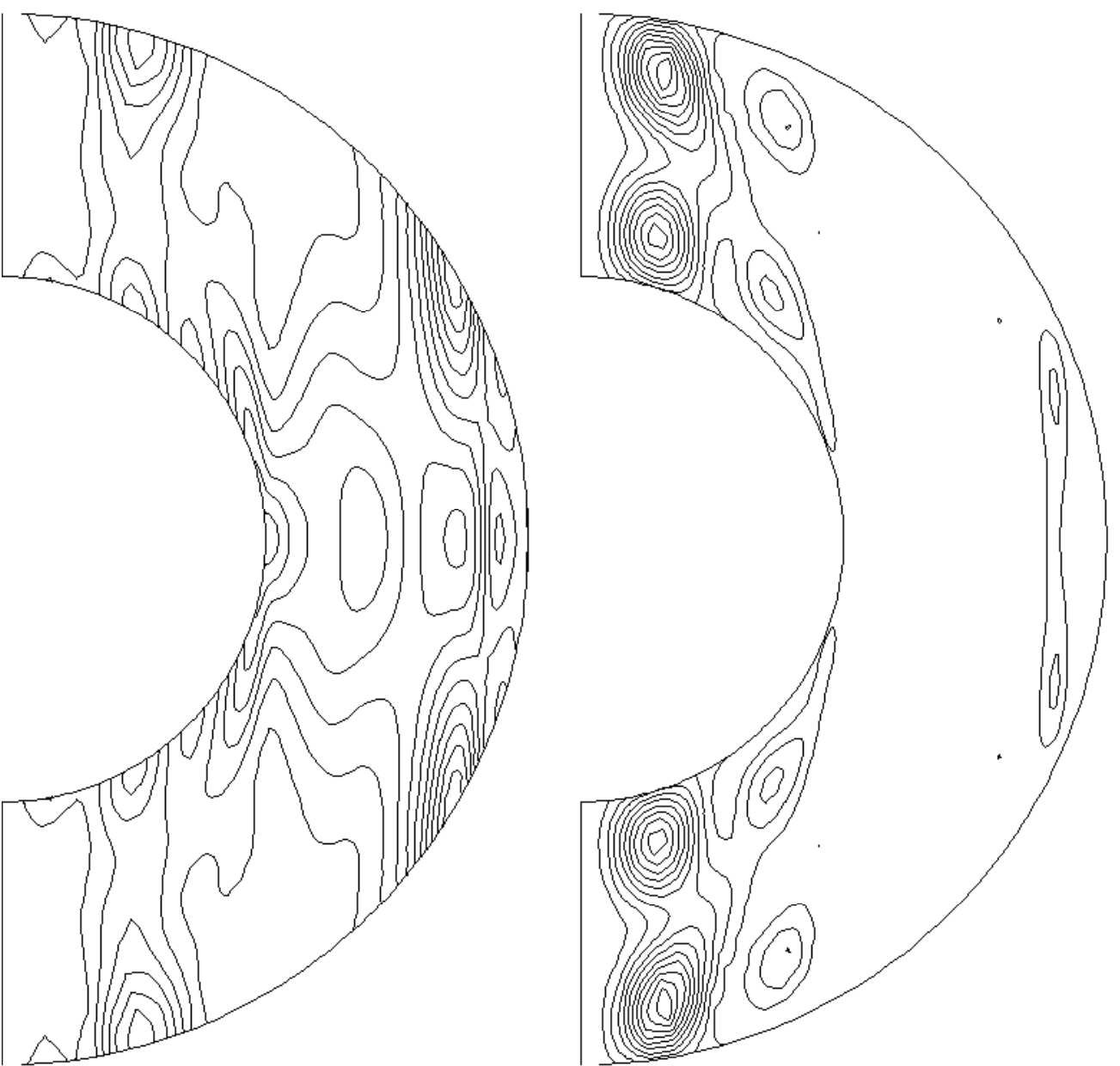}\label{fig3b}}
\subfigure[]{\includegraphics[scale=0.4]{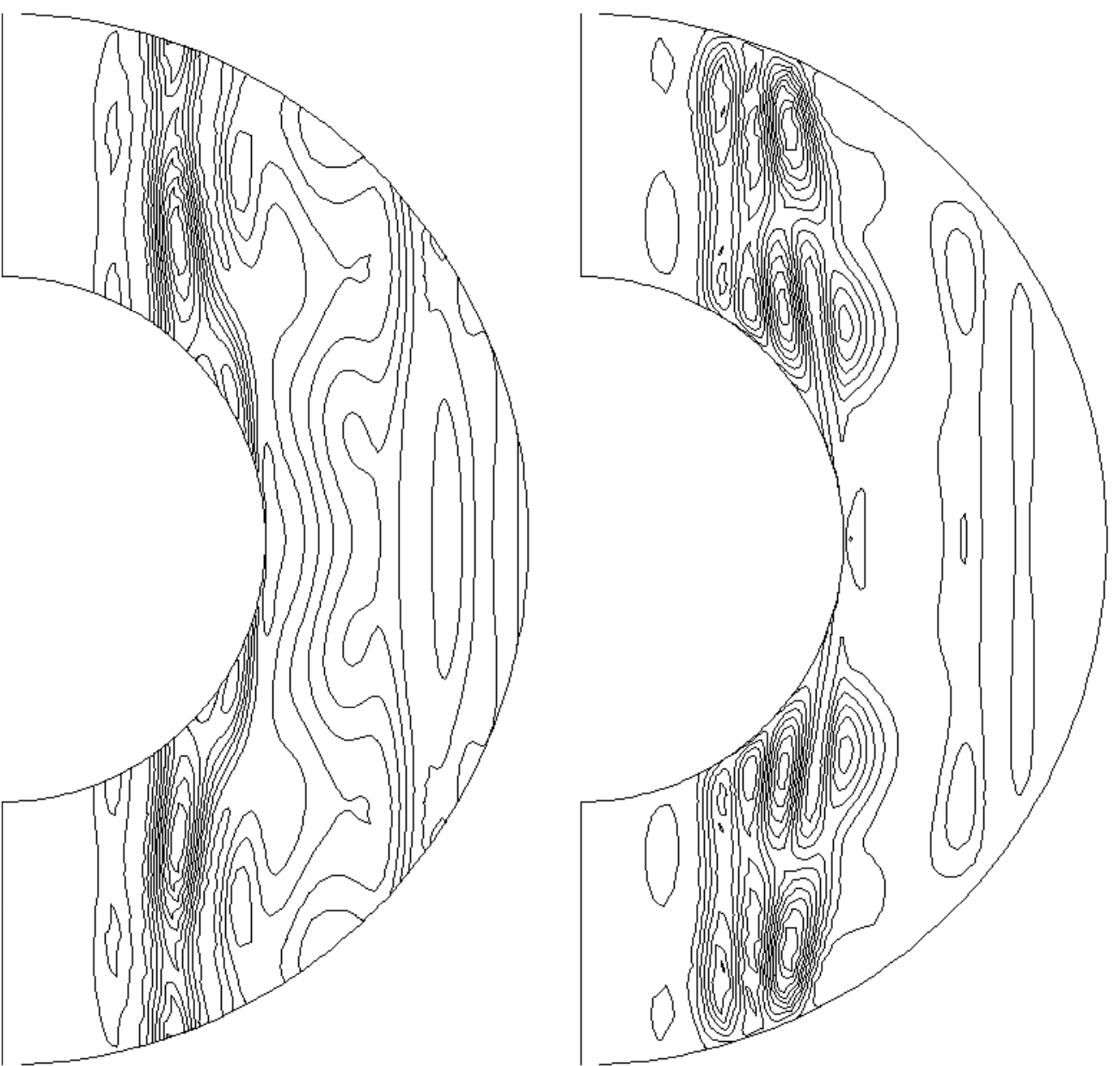}\label{fig3c}}
\subfigure[]{\includegraphics[scale=0.4]{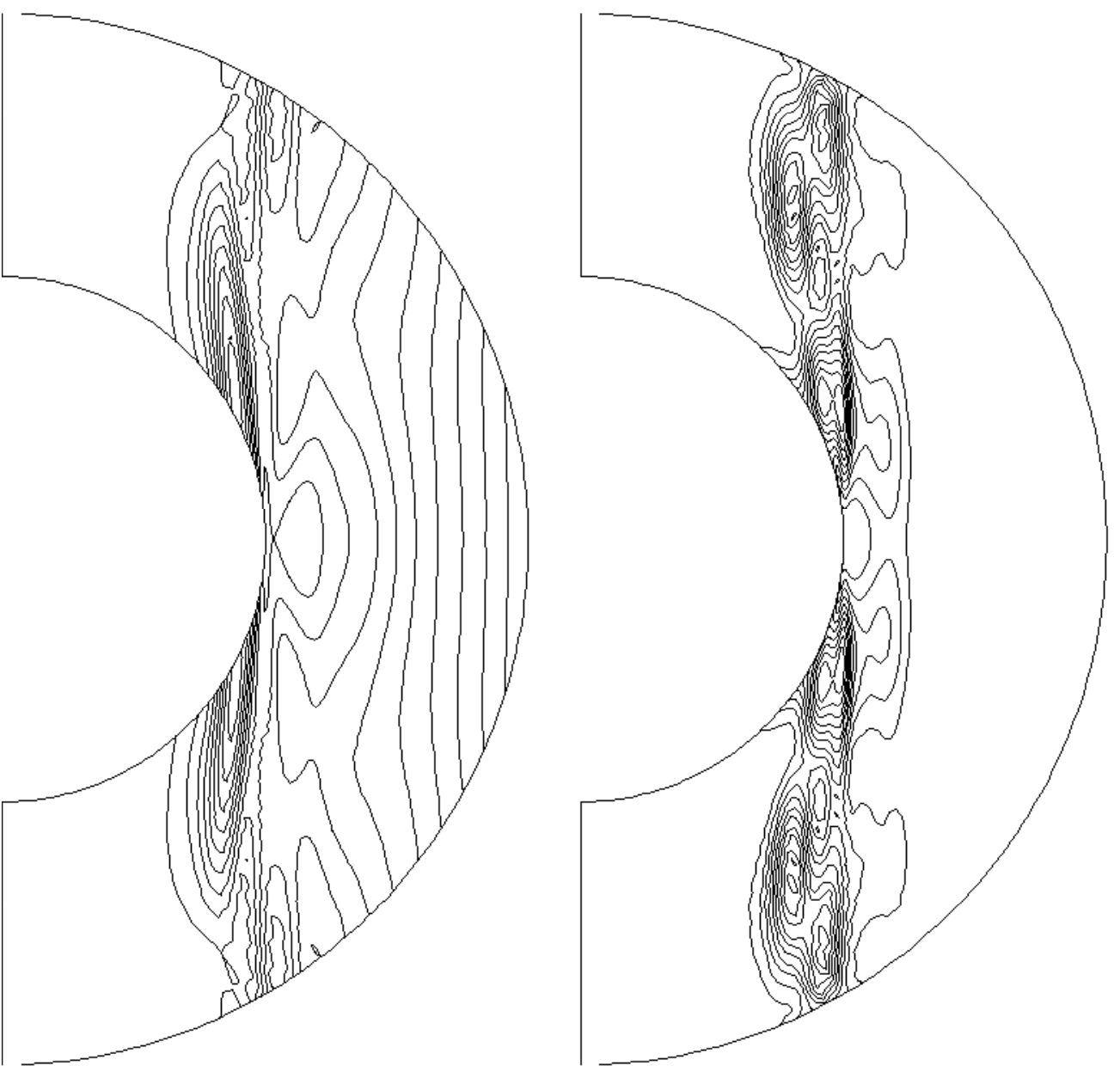}\label{fig3d}}
\subfigure[]{\includegraphics[scale=0.4]{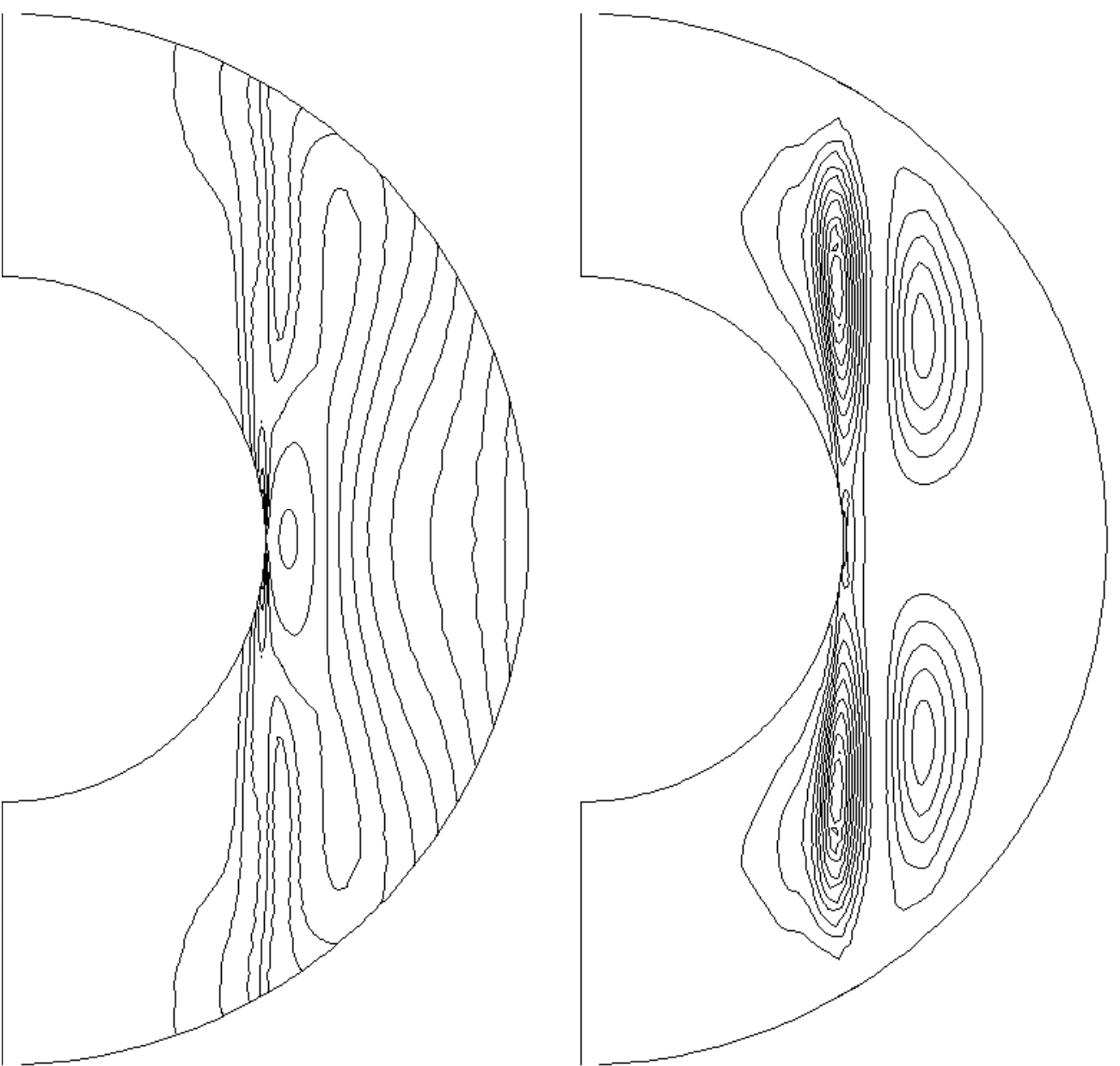}\label{fig3e}}
\subfigure[]{\includegraphics[scale=0.4]{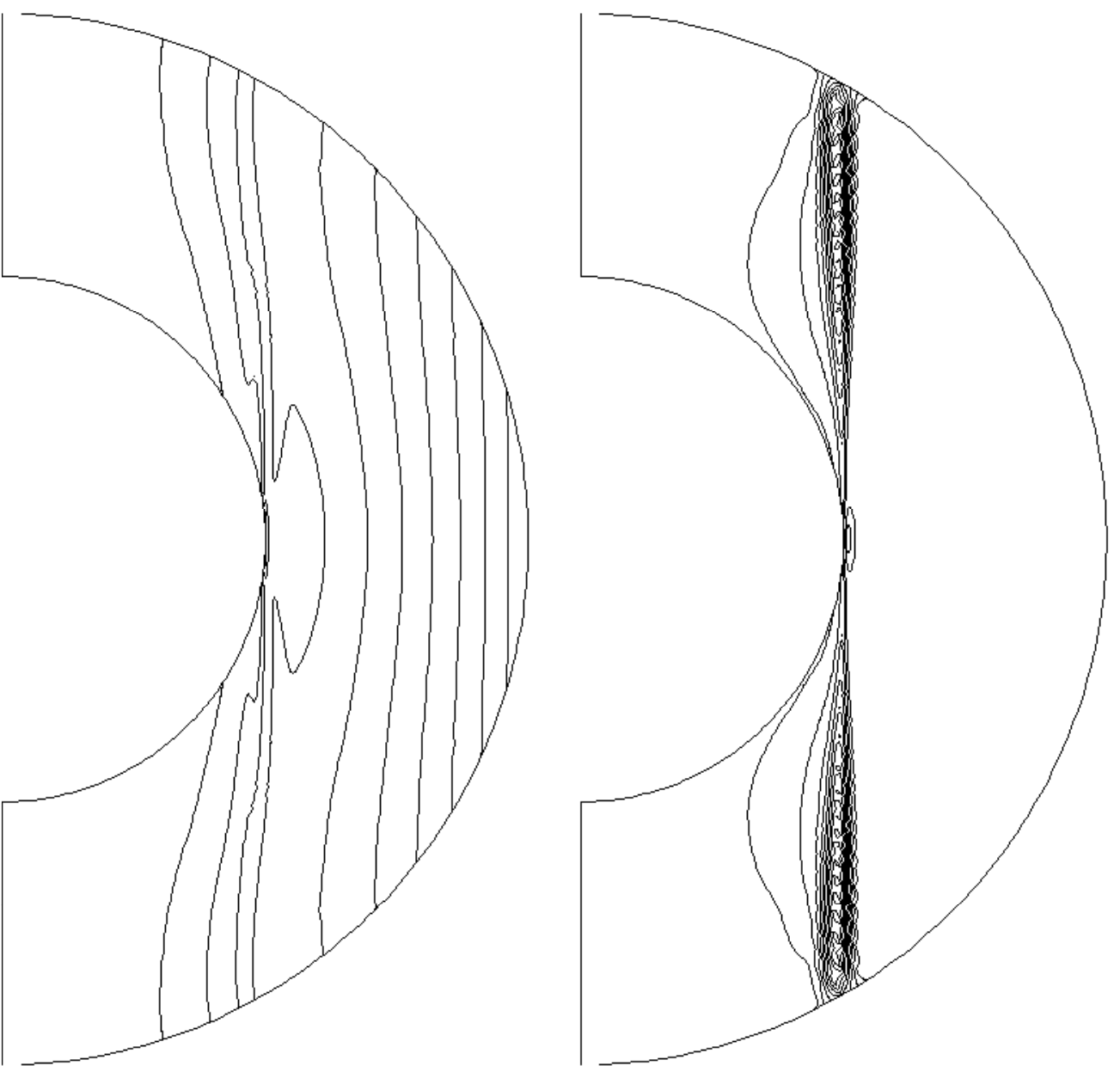}\label{fig3f}}
\caption{(a) shows the contours of kinetic energy at $S=0$, and (b)-(f) show the contours of kinetic (left panel) and magnetic (right panel) energies at, respectively, $S^2=0.01$, $0.02$, $0.05$, $0.1$ and $0.2$. The tidal frequency is $\omega=1.0$.}\label{fig3}
\end{figure}

To better understand the mechanism of tidal dissipations with magnetic field, we plot the contours of kinetic and magnetic energies in a certain meridional plane at $\phi=90^\circ$. Here magnetic energy is in terms of the induced field, but does not include the contribution of imposed field. Figure \ref{fig3} shows the energy contours at the tidal frequency $\omega=1.0$. Figure \ref{fig3a} is for rotating flow and Figures \ref{fig3b}-\ref{fig3f} are for rotating MHD with the field strength gradually increasing. In Figure \ref{fig3} both tidal flow (equilibrium tide) and waves (dynamical tide) are included. Inertial waves in rotating flow propagate at an angle $\arcsin(\omega/2\Omega)=30^\circ$ inclined to the rotational axis and Figure \ref{fig3a} exhibits the $30^\circ$ oblique internal shear layers built by the propagation of inertial waves. However, in the presence of magnetic field as shown in Figures \ref{fig3b}-\ref{fig3f}, these oblique internal shear layers disappear. With the field strength gradually increasing, the contours of kinetic energy become more vertical and those of magnetic energy concentrate on the cylinder tangent to the inner sphere. This is because the magneto-inertial waves become more Alfv\'en-like than inertial-like when the field strength increases. In the meanwhile, the kinetic energy spreads outside the tangent cylinder whereas the magnetic energy concentrates more in a thin internal shear layer on the tangent cylinder. This difference between the flow and field structures explains why magnetic dissipation wins out viscous dissipation in the low-frequency range with a strong field because the smaller length scale of magnetic field leads to the higher magnetic dissipation. It should be noted that with a very weak magnetic field, e.g. $S=10^{-4}$, the magnetic effect is negligible and the internal shear layers cannot be destroyed by the field \citep{ogilvie2017}. So magnetic field wins out rotation at a moderate or stronger field in the sense that the Alfv\'en velocity is at least of the order of 0.1 of the surface rotational velocity. A more accurate scaling law for this competition was obtained by \citet{ogilvie2017}.

\begin{figure}
\centering
\subfigure[]{\includegraphics[scale=0.4]{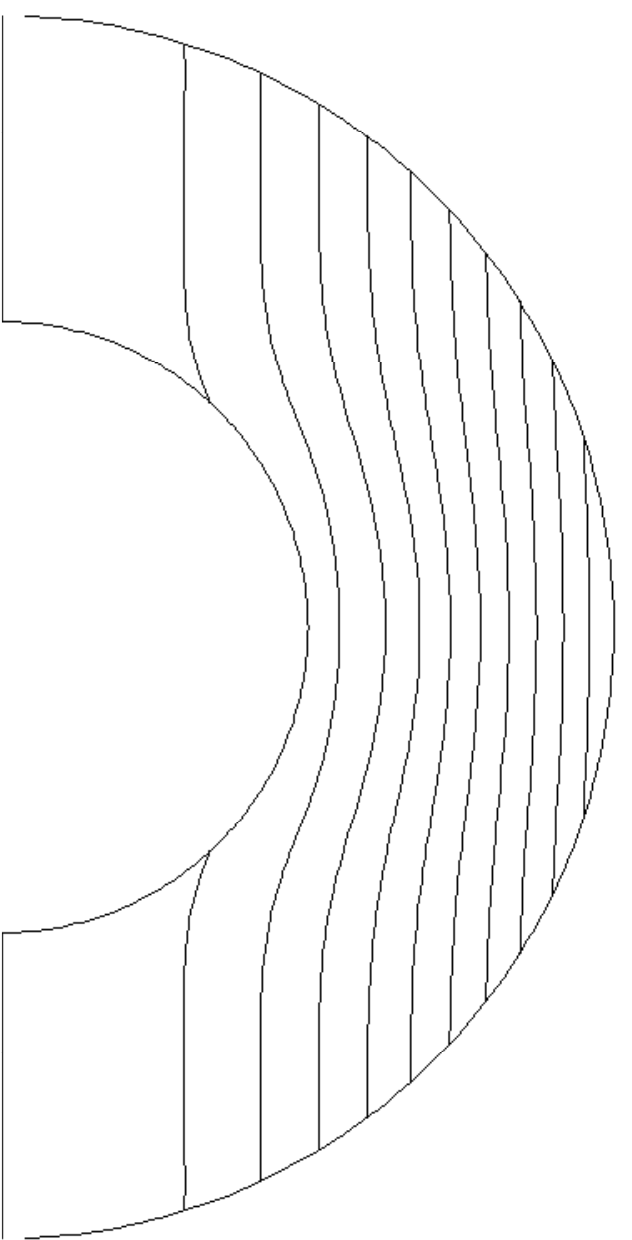}\label{fig4a}} \\
\subfigure[]{\includegraphics[scale=0.4]{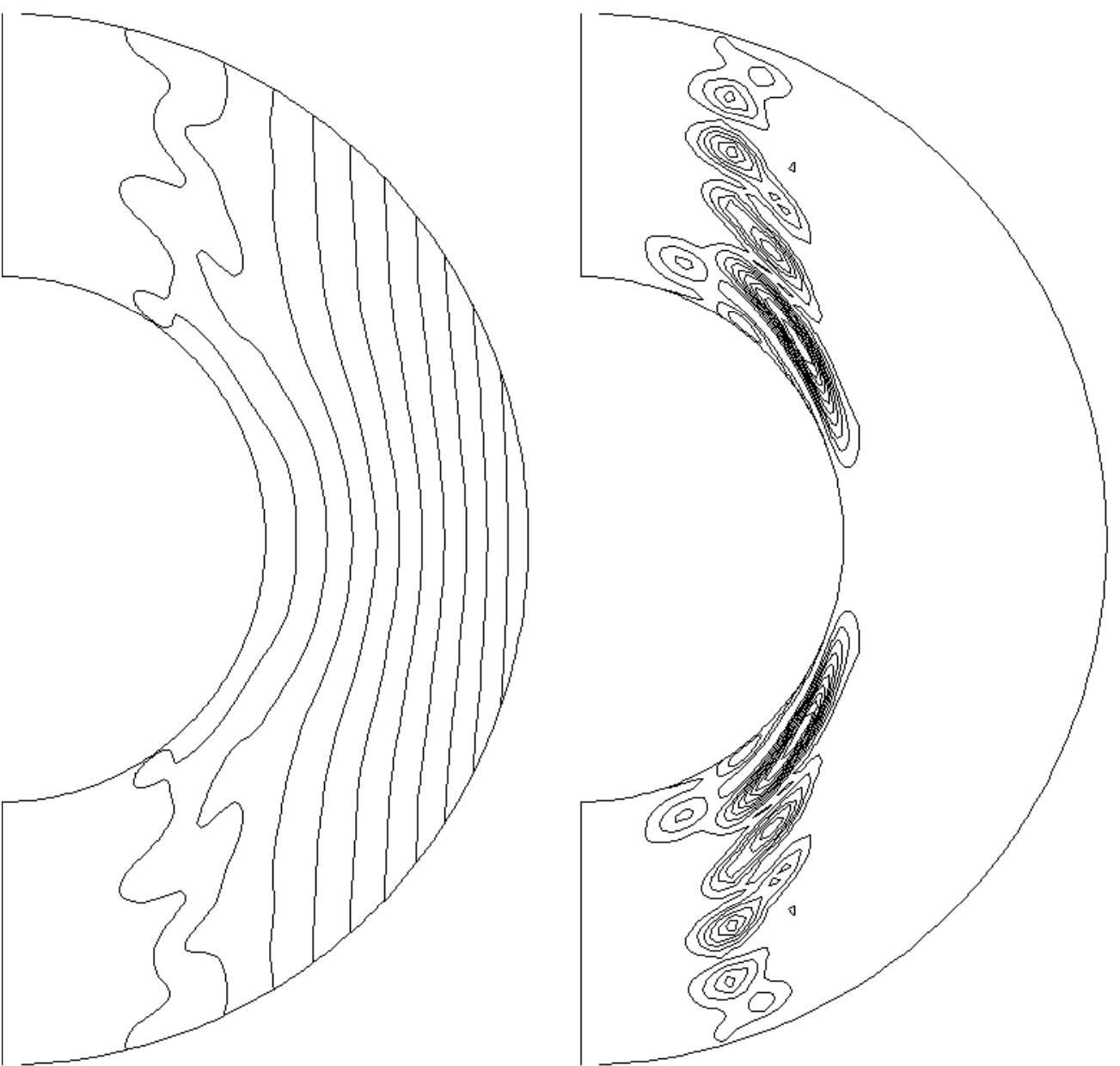}\label{fig4b}}
\subfigure[]{\includegraphics[scale=0.4]{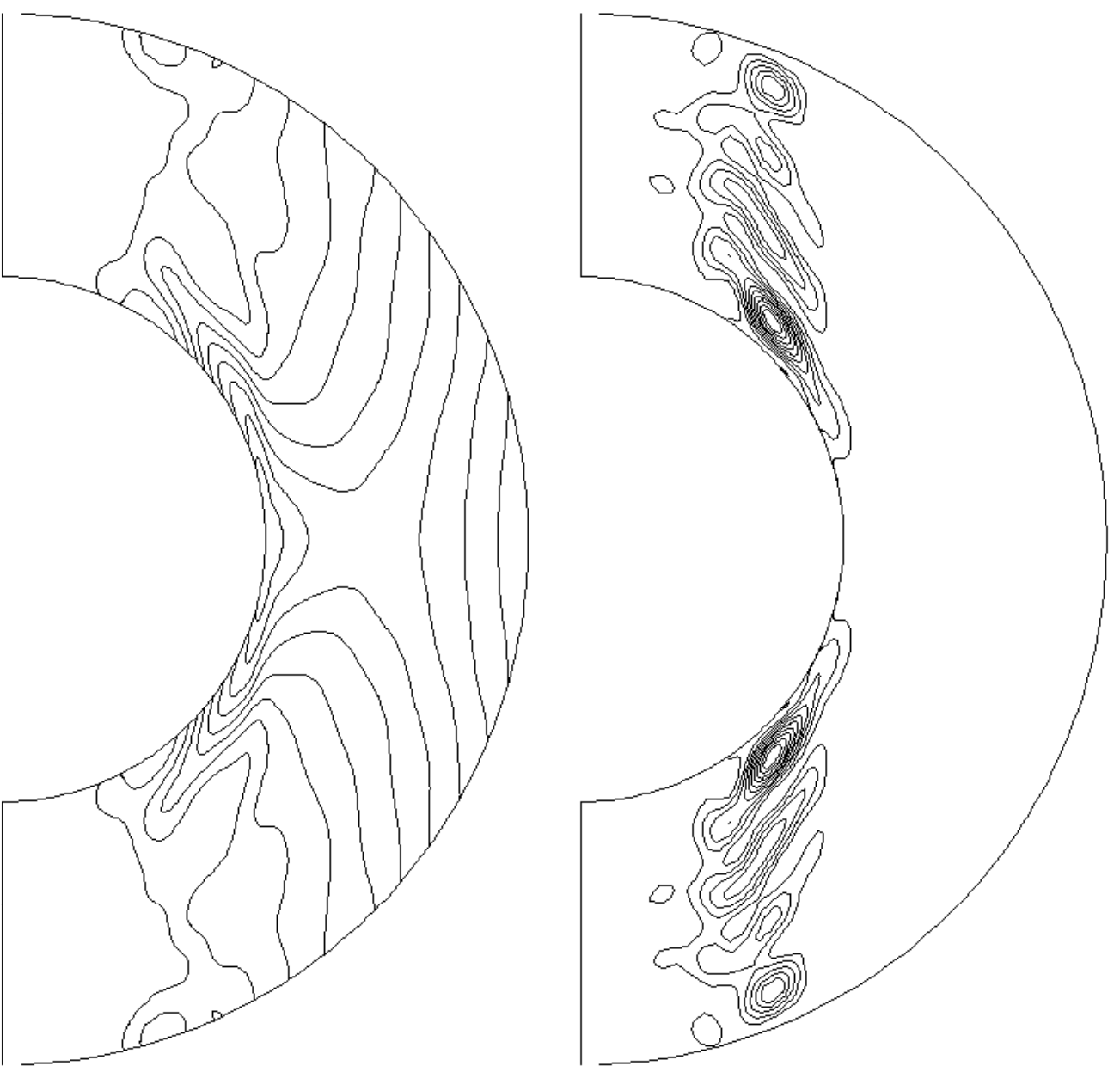}\label{fig4c}}
\subfigure[]{\includegraphics[scale=0.4]{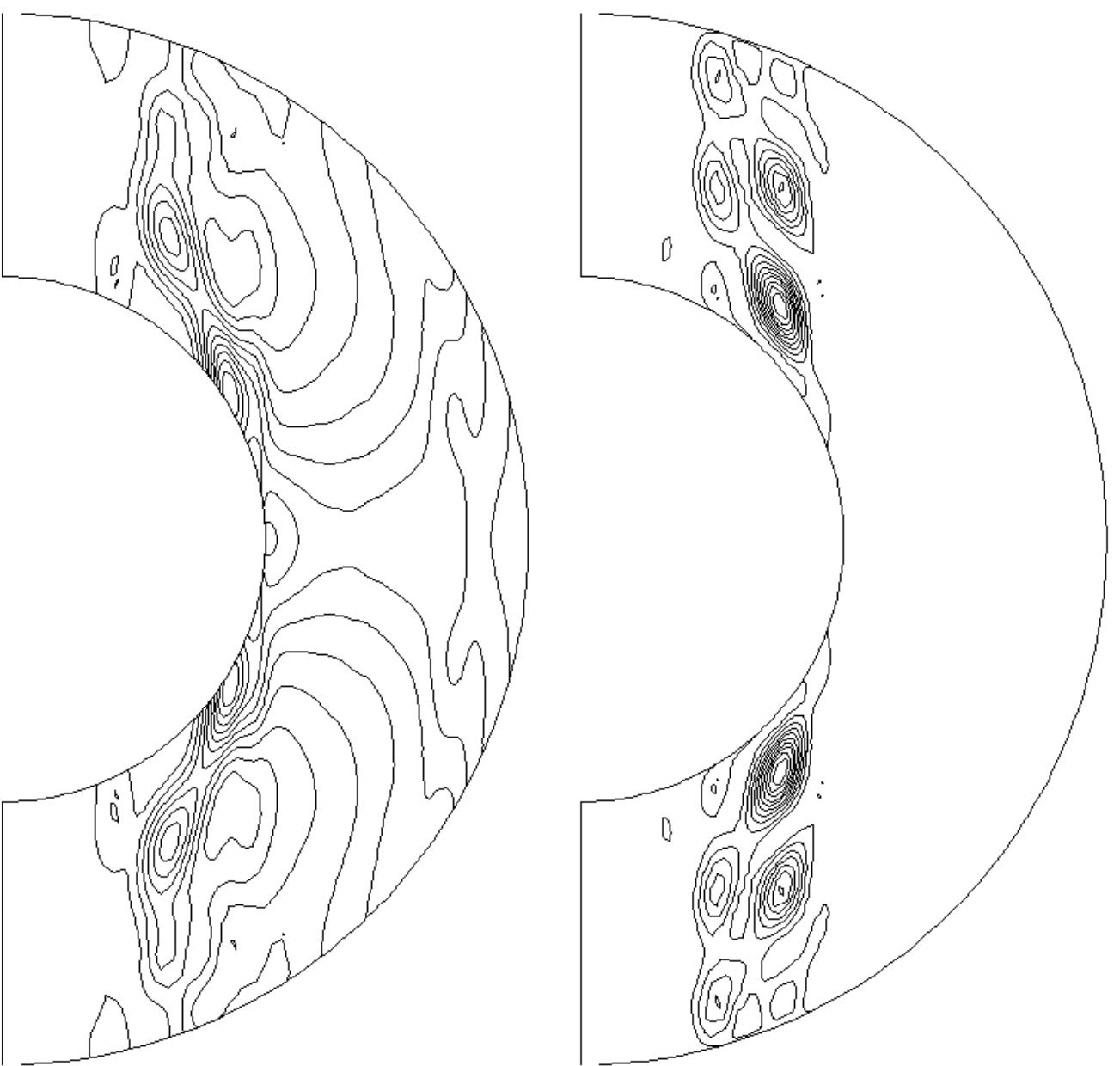}\label{fig4d}}
\subfigure[]{\includegraphics[scale=0.4]{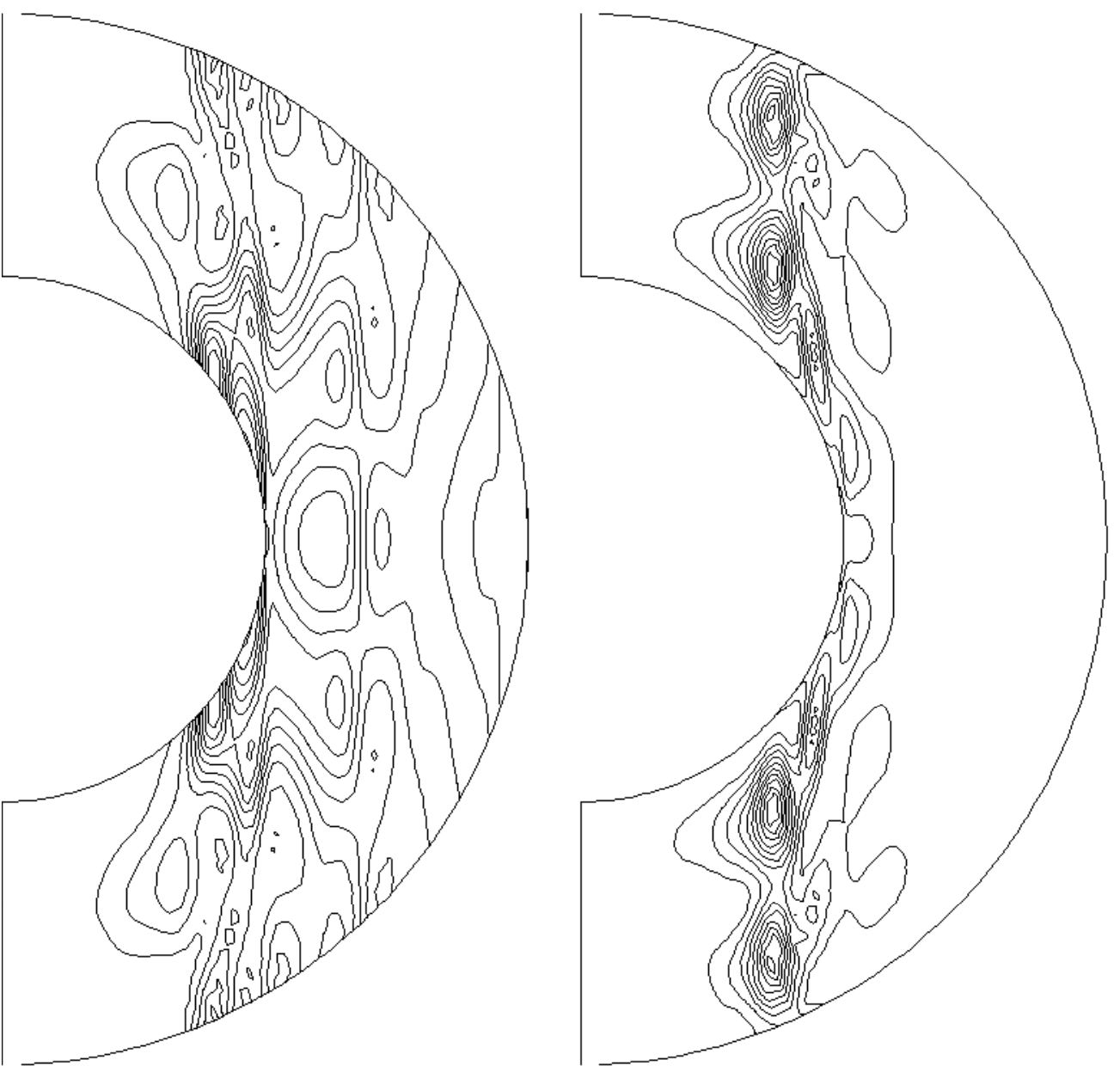}\label{fig4e}}
\subfigure[]{\includegraphics[scale=0.4]{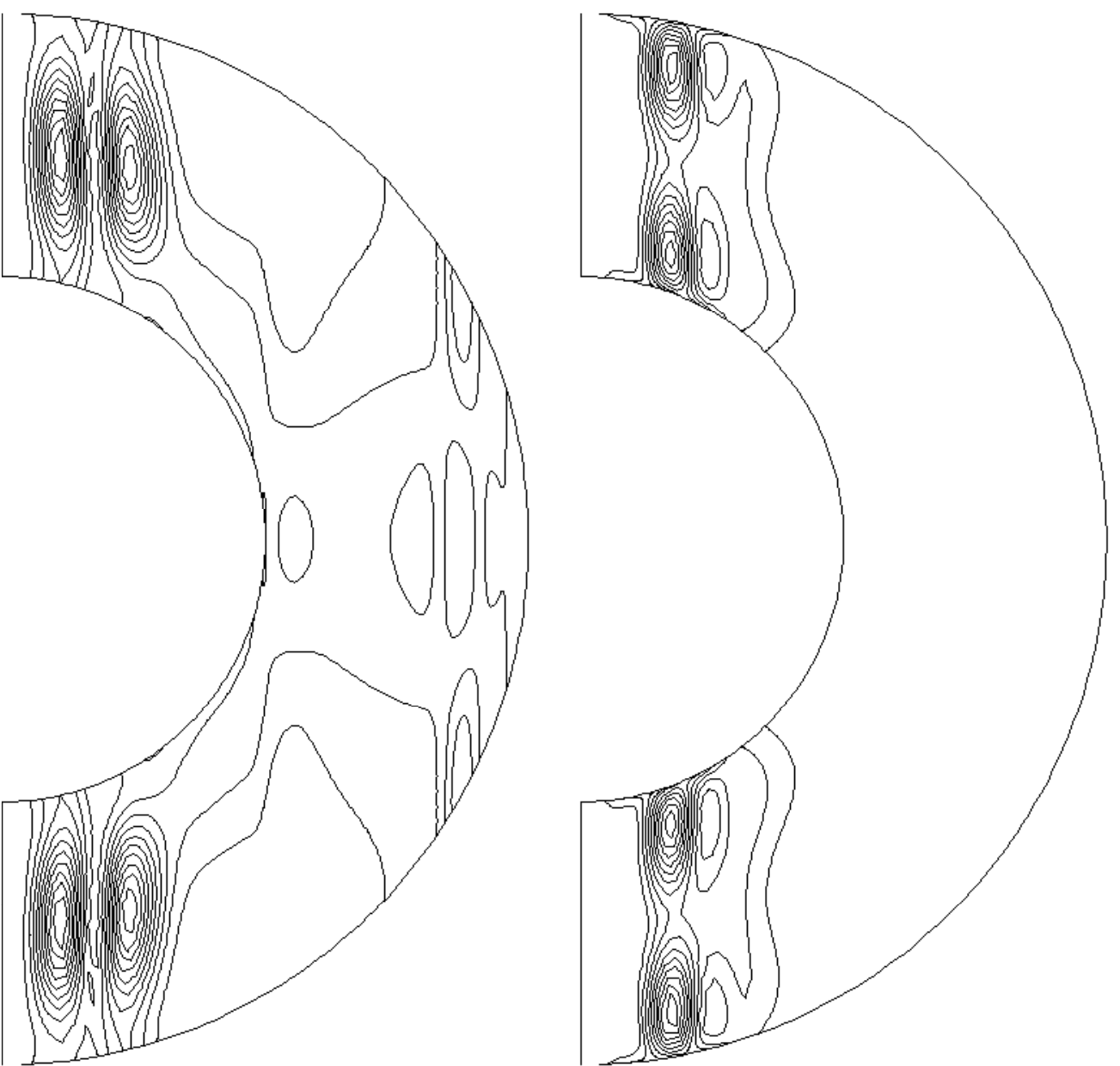}\label{fig4f}}
\caption{The same as Figure \ref{fig3} but for the tidal frequency $\omega=3.0$.}\label{fig4}
\end{figure}

For comparison with the low tidal frequency, we plot the energy contours at a high tidal frequency $\omega=3.0$, as shown in Figure \ref{fig4}. Figure \ref{fig4a} shows that the contours of kinetic energy of rotating flow consisting of both equilibrium tide and dynamical tide at the high tidal frequency are almost vertical but do not exhibit the structure of oblique internal shear layers at the low tidal frequency because the tidal frequency higher than $2\Omega$ cannot excite inertial waves to build the oblique internal shear layers. When magnetic field is present as shown in Figures \ref{fig4b}-\ref{fig4f}, both kinetic and magnetic energies exhibit similar distributions and tend to concentrate inside the tangent cylinder when the field strength increases, which explains why viscous and magnetic dissipations are comparable at the high tidal frequency.

\begin{figure}
\centering
\subfigure[]{\includegraphics[scale=0.45]{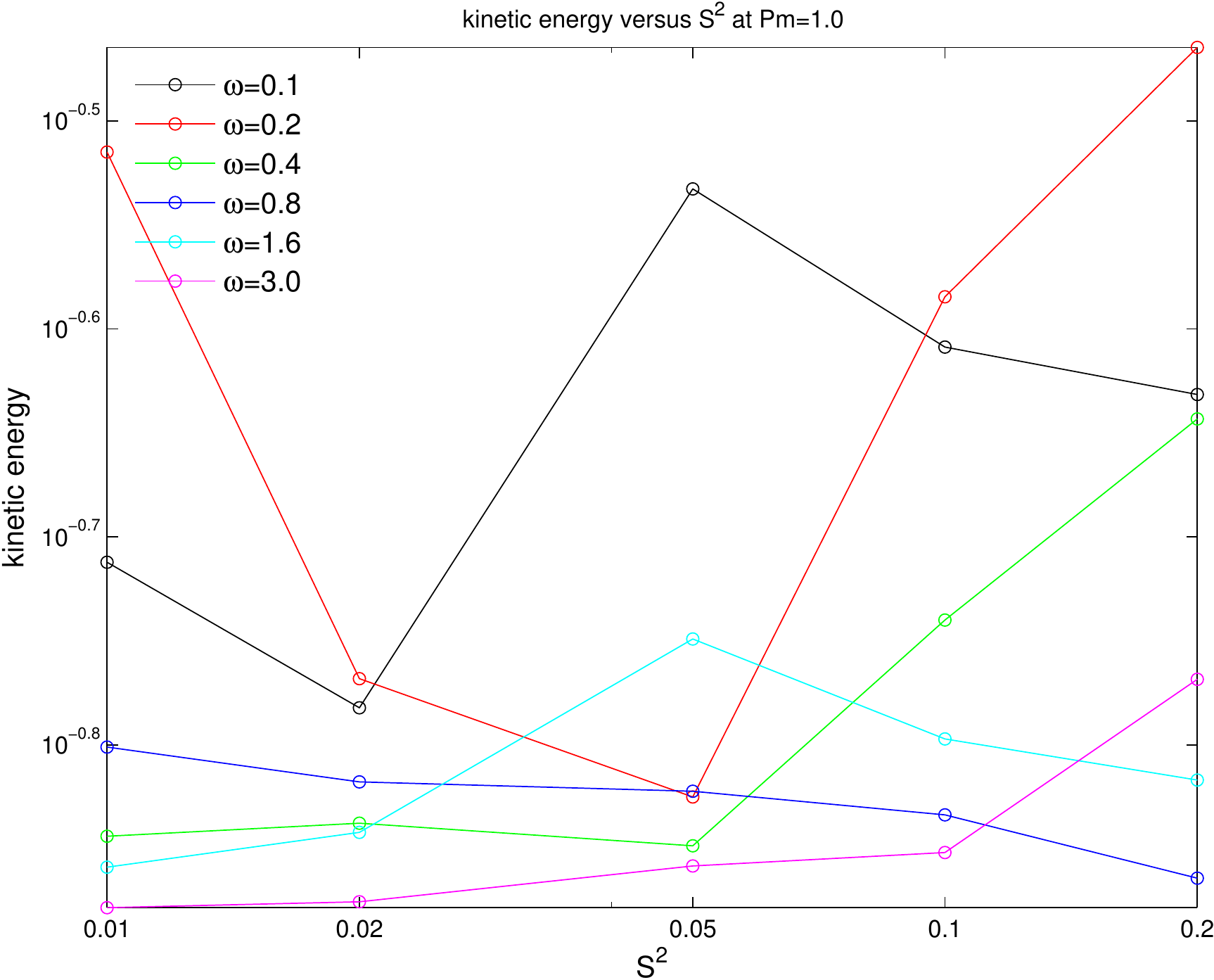}\label{fig5a}} \\
\subfigure[]{\includegraphics[scale=0.45]{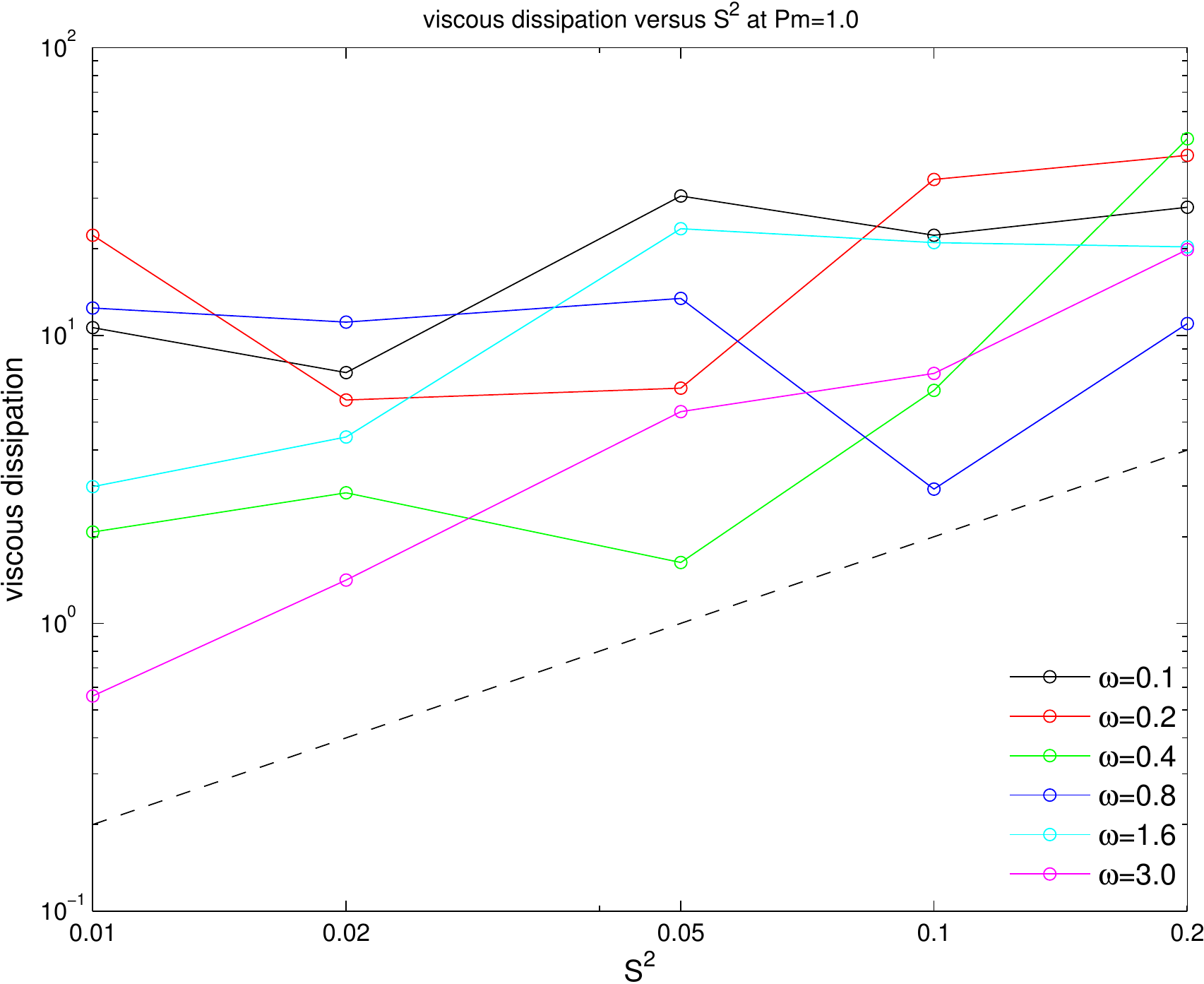}\label{fig5b}}
\subfigure[]{\includegraphics[scale=0.45]{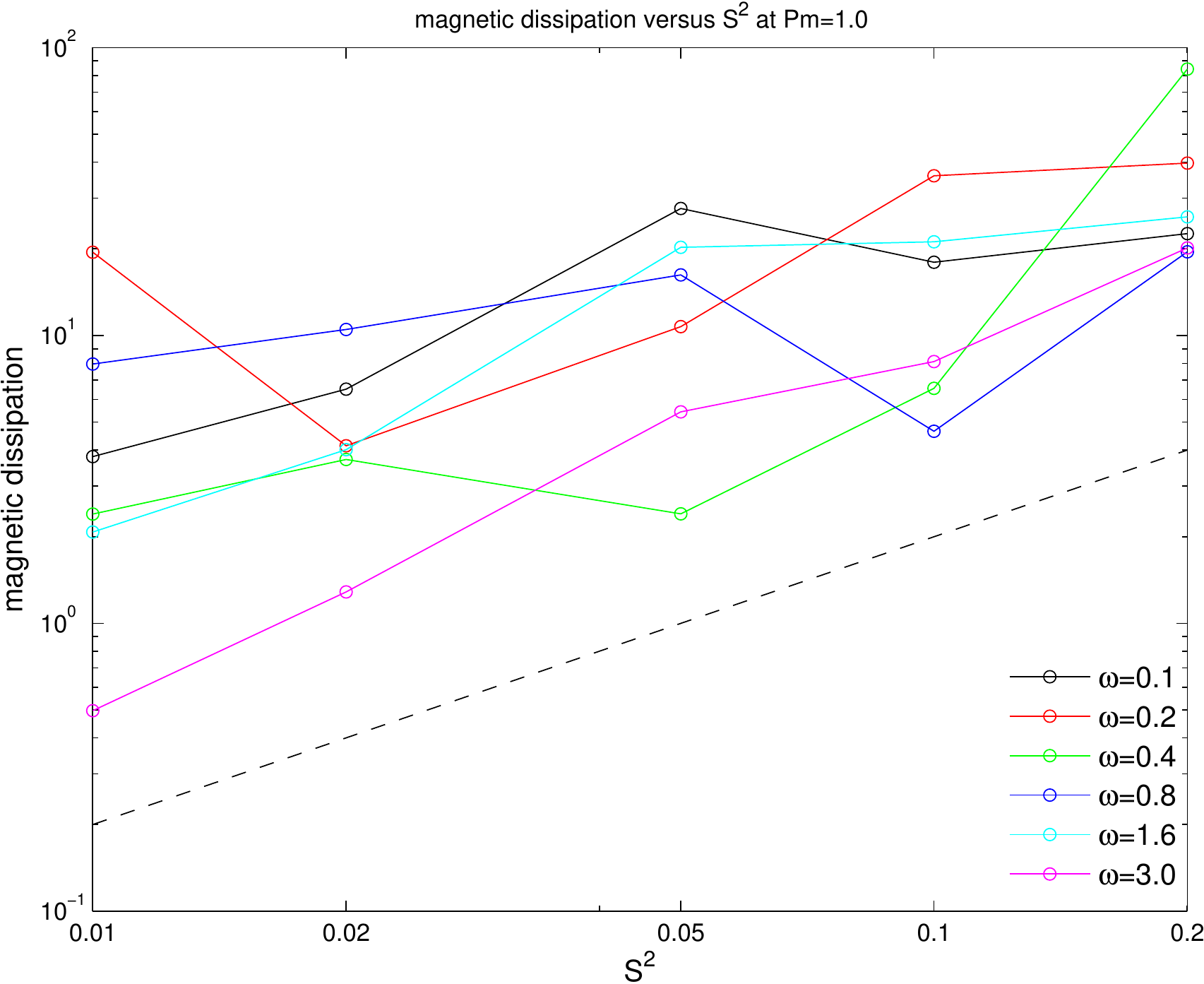}\label{fig5c}}
\caption{(a) kinetic energy (b) viscous dissipation and (c) magnetic dissipation versus $S^2$ at various $\omega$. $Pm=1.0$. The two dashed lines in \ref{fig5b} and \ref{fig5c} denote the scaling law $\propto S^2$.}\label{fig5}
\end{figure}

The dependence of dissipations on tidal frequency is irregular because the resonance in spherical geometry is complicated, but the dependence of dissipations on magnetic field is not so irregular. Figure \ref{fig5} shows the log-log relation of kinetic energy and dissipations versus $S^2$ at the different tidal frequencies which are almost a geometric sequence. Figure \ref{fig5a} shows the kinetic energy versus $S^2$. It is irregular because the resonance occurs irregularly at certain frequencies with certain field strengths. For dissipations shown in Figures \ref{fig5b} and \ref{fig5c}, we cannot find accurate scaling laws, i.e. straight lines in the log-log diagram, because certain tidal frequencies are near resonance at certain $S$ values as shown in Figure \ref{fig5a}. However, we can perceive that both viscous and magnetic dissipations increase with the field strength increasing. It is not surprising because stronger field provides higher magnetic energy to damp. Then we fit the slopes of this `big trend' for all the curves by appropriately removing few points far away from this `big trend' and take the average of these slopes. Interestingly, both viscous and magnetic dissipations obey an identical scaling law, 
\begin{equation}\label{eq:S-scaling}
\mbox{viscous and magnetic dissipations} \propto S^2.
\end{equation}
The dashed lines in Figures \ref{fig5b} and \ref{fig5c} denote this scaling law. This scaling law about the relation of the two dissipations and magnetic field reveals that there exists self-similarity for the magnetic effect on dynamical tide.

\subsection{Investigation of $Pm$}

\begin{figure}
\centering
\subfigure[]{\includegraphics[scale=0.45]{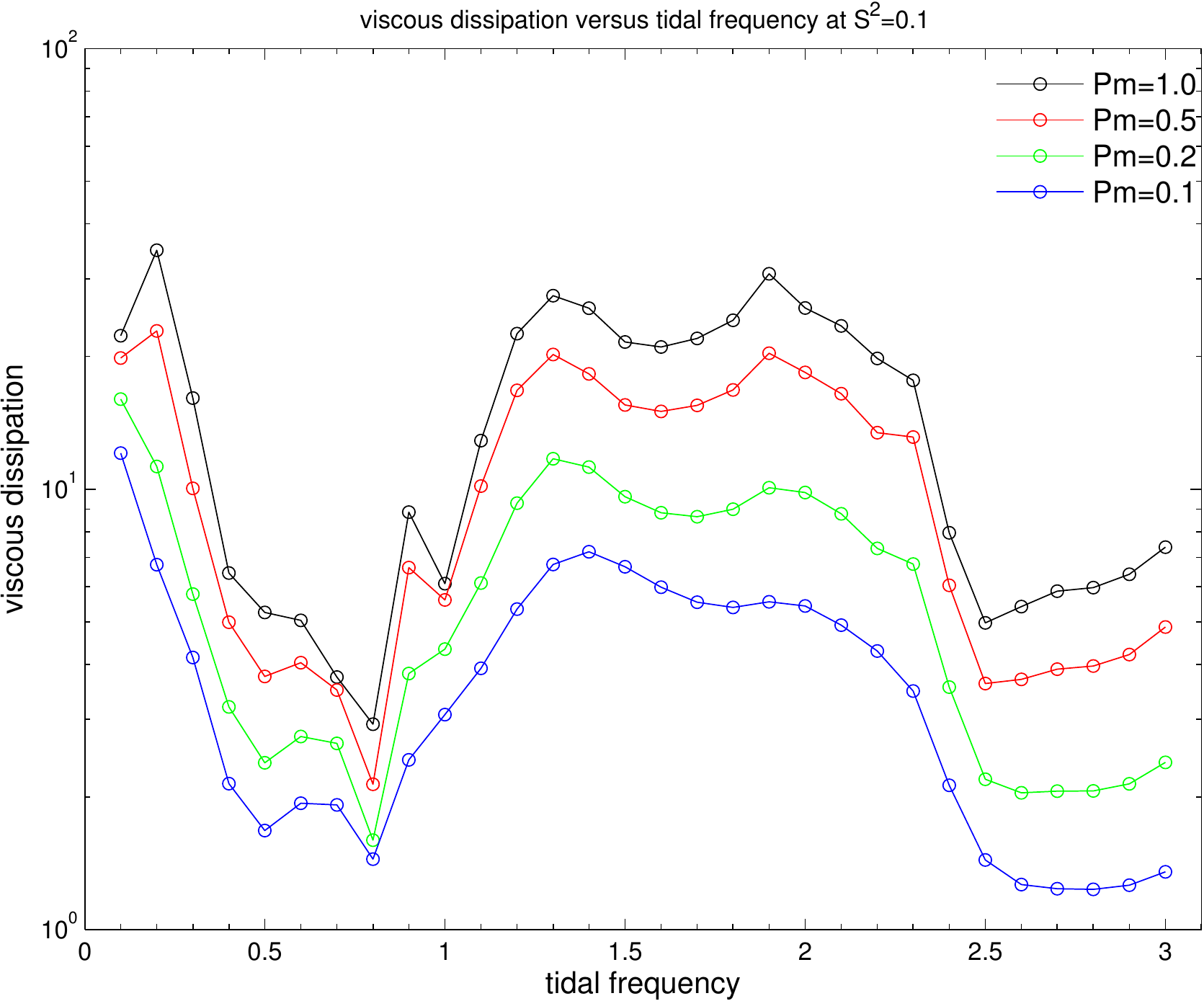}\label{fig6a}}
\subfigure[]{\includegraphics[scale=0.45]{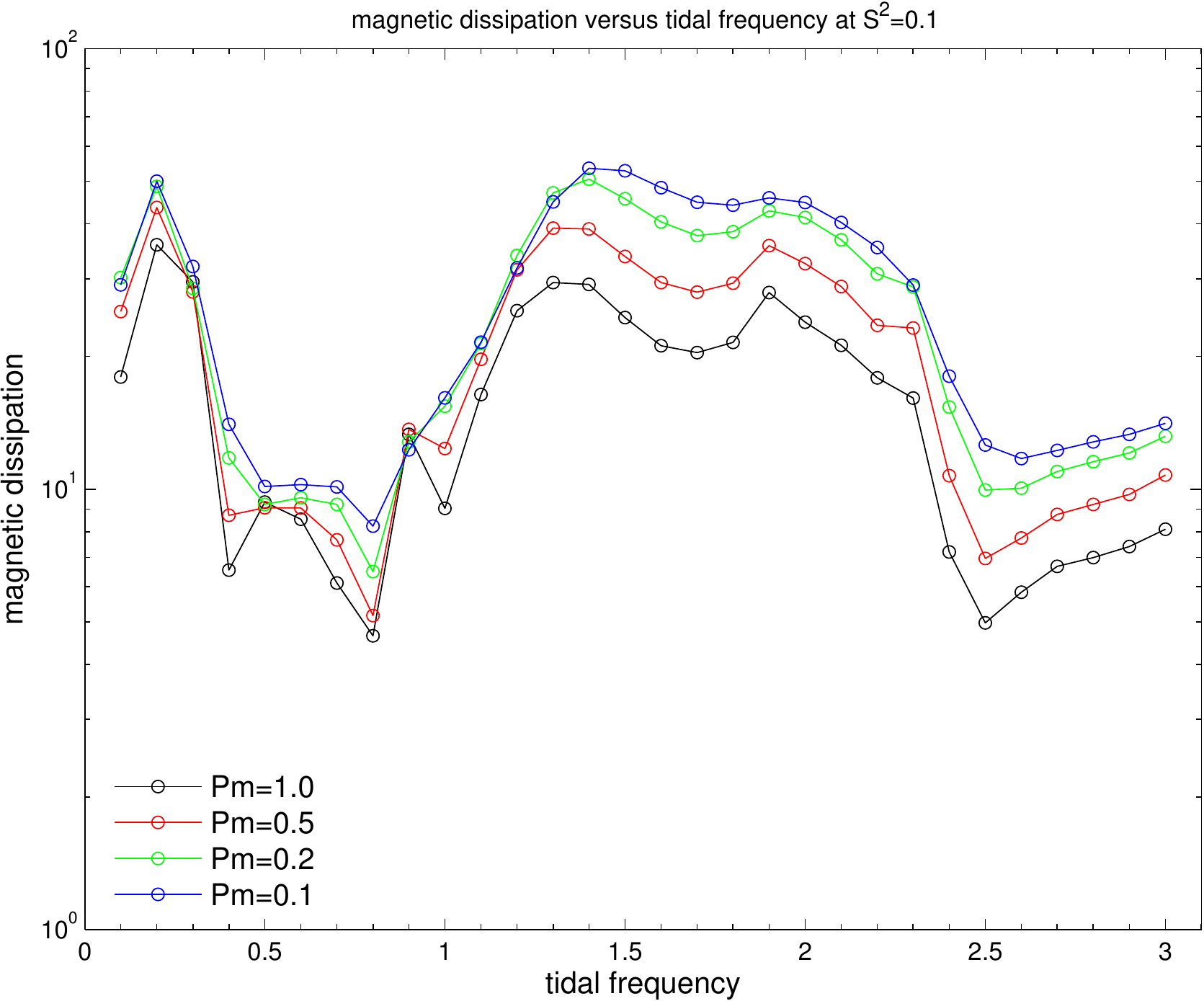}\label{fig6b}}
\subfigure[]{\includegraphics[scale=0.45]{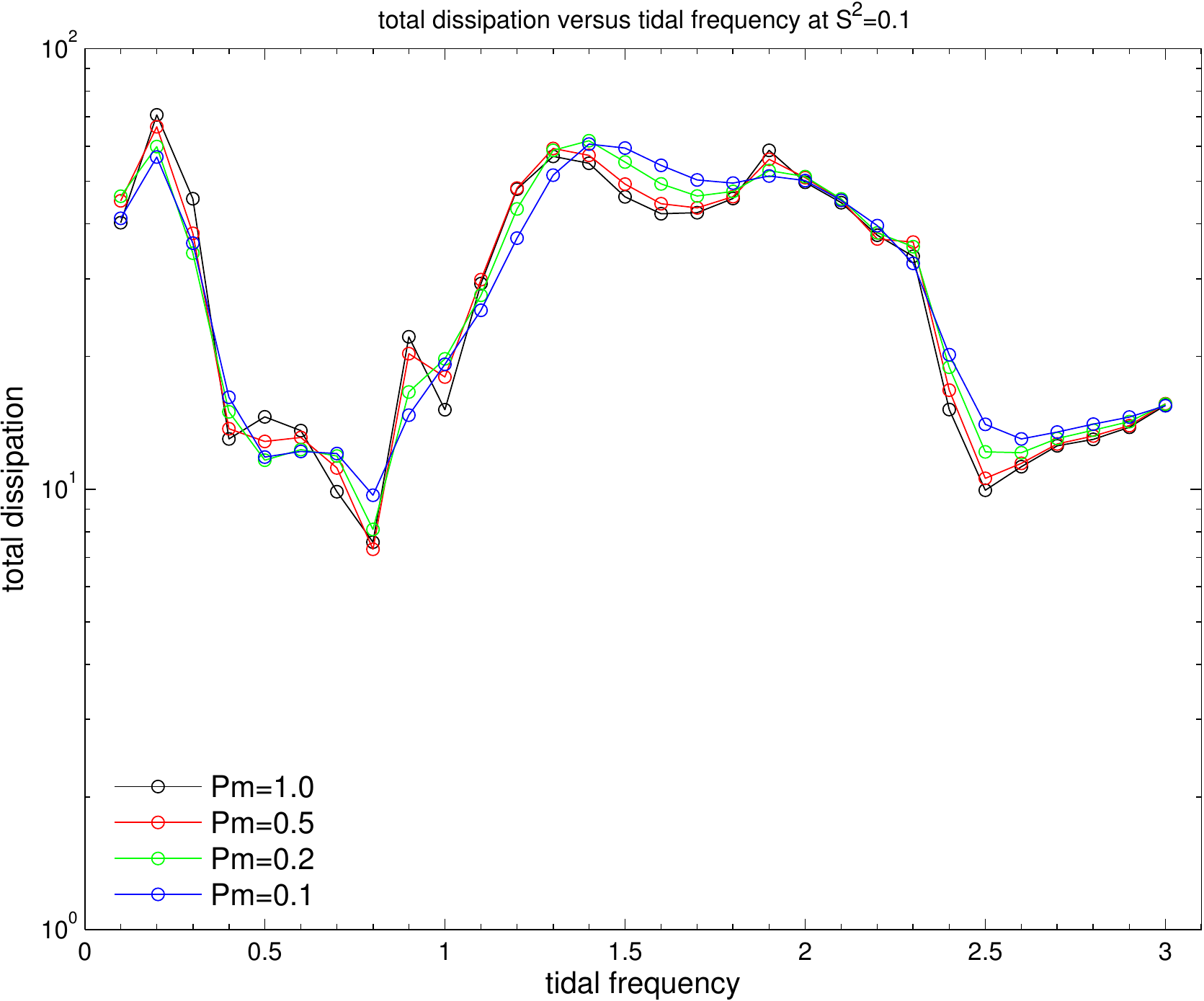}\label{fig6c}}
\subfigure[]{\includegraphics[scale=0.45]{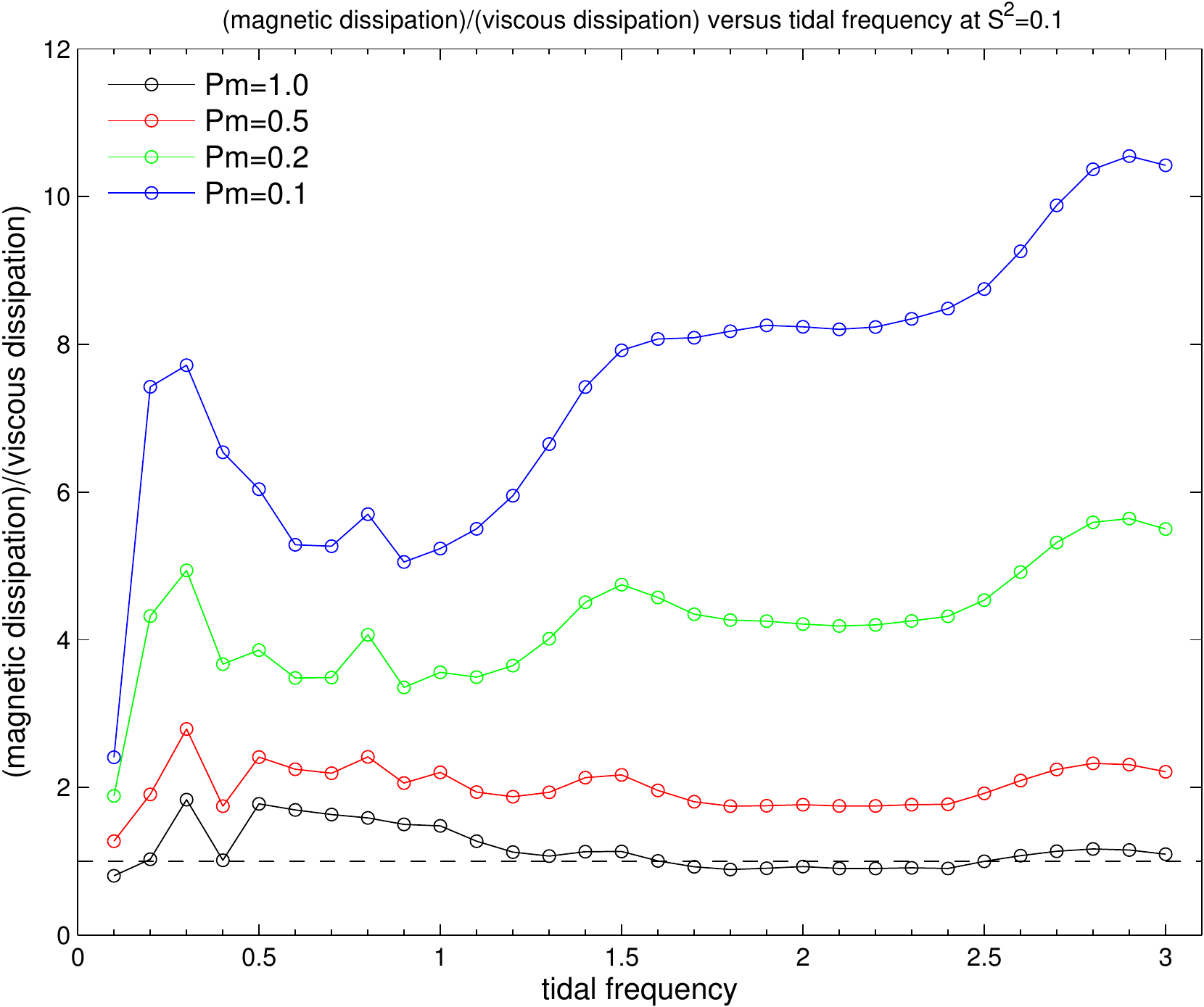}\label{fig6d}}
\caption{(a) Viscous dissipation, (b) magnetic dissipation, (c) total dissipation and (d) ratio of magnetic to viscous dissipations versus tidal frequency at various $Pm$. $S^2=0.1$.}\label{fig6}
\end{figure}

In this subsection we investigate $Pm$. Similar to Figure \ref{fig2}, Figure \ref{fig6} shows viscous dissipation, magnetic dissipation, total dissipation and ratio of magnetic to viscous dissipations versus tidal frequency at various $Pm$ with $S^2$ fixed to be 0.1, namely in a strong-field regime. The first three subfigures show that for all the tidal frequencies a larger $Pm$ leads to higher viscous dissipation and lower magnetic dissipation but total dissipation is almost independent of $Pm$. This is reasonable because $Pm$ is the ratio of viscosity to magnetic diffusivity (see Equation \eqref{eq:pm}) but cannot influence the total diffusivity. More interestingly, Figure \ref{fig6d} shows that magnetic dissipation wins out viscous dissipation for $Pm<1$ and this ratio is approximately equal to $Pm^{-1}$. In the astrophysical situation $Pm$ is very small, and it can be inferred that {\bf magnetic dissipation dominates over viscous dissipation with a moderate or stronger magnetic field}. It should be noted again that with a very weak field, e.g. $S\le10^{-4}$, viscous dissipation wins out magnetic dissipation even at a small $Pm$, e.g. $Pm=10^{-4}$ \citep{ogilvie2017}.

\begin{figure}
\centering
\subfigure[]{\includegraphics[scale=0.45]{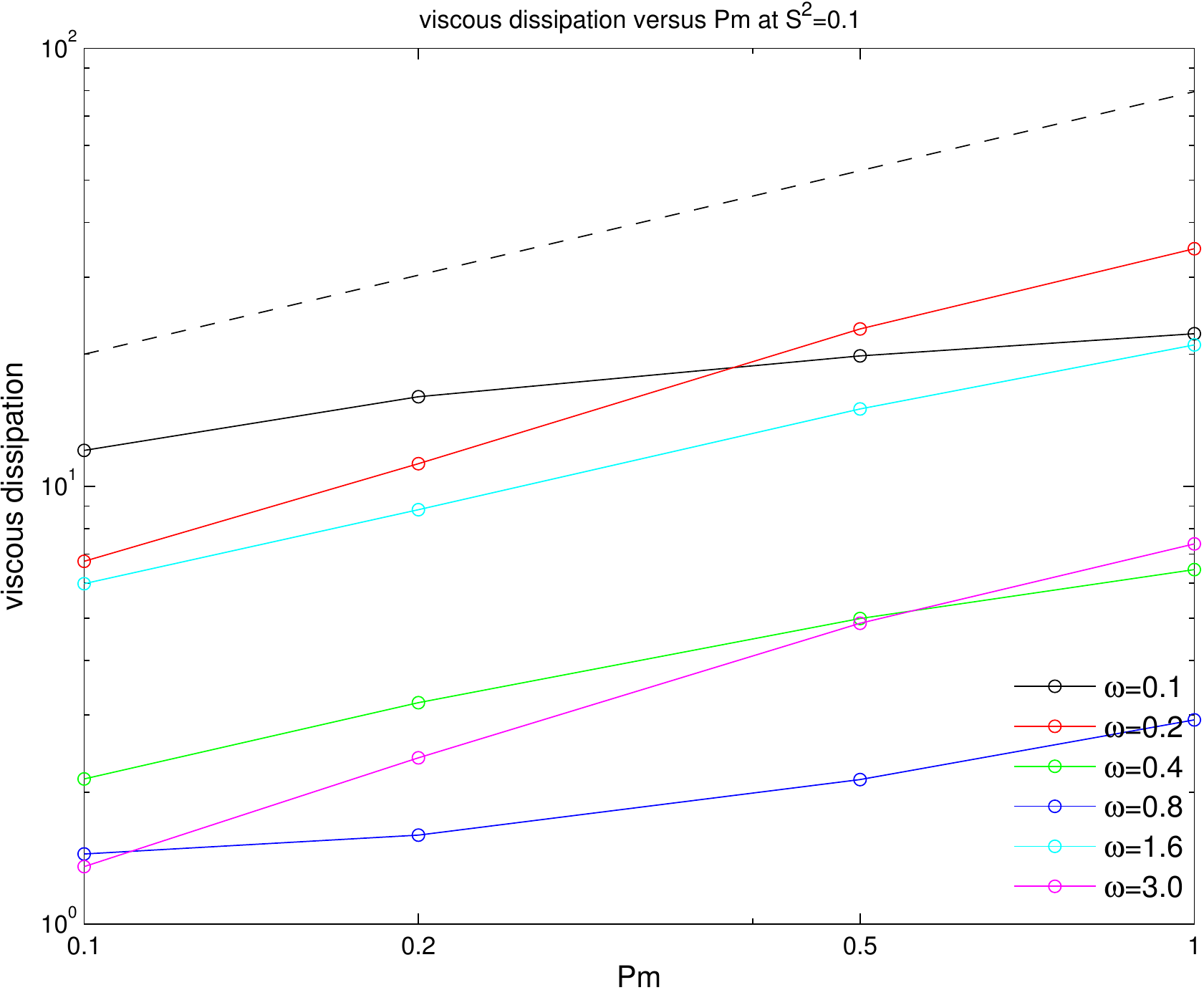}\label{fig7a}}
\subfigure[]{\includegraphics[scale=0.45]{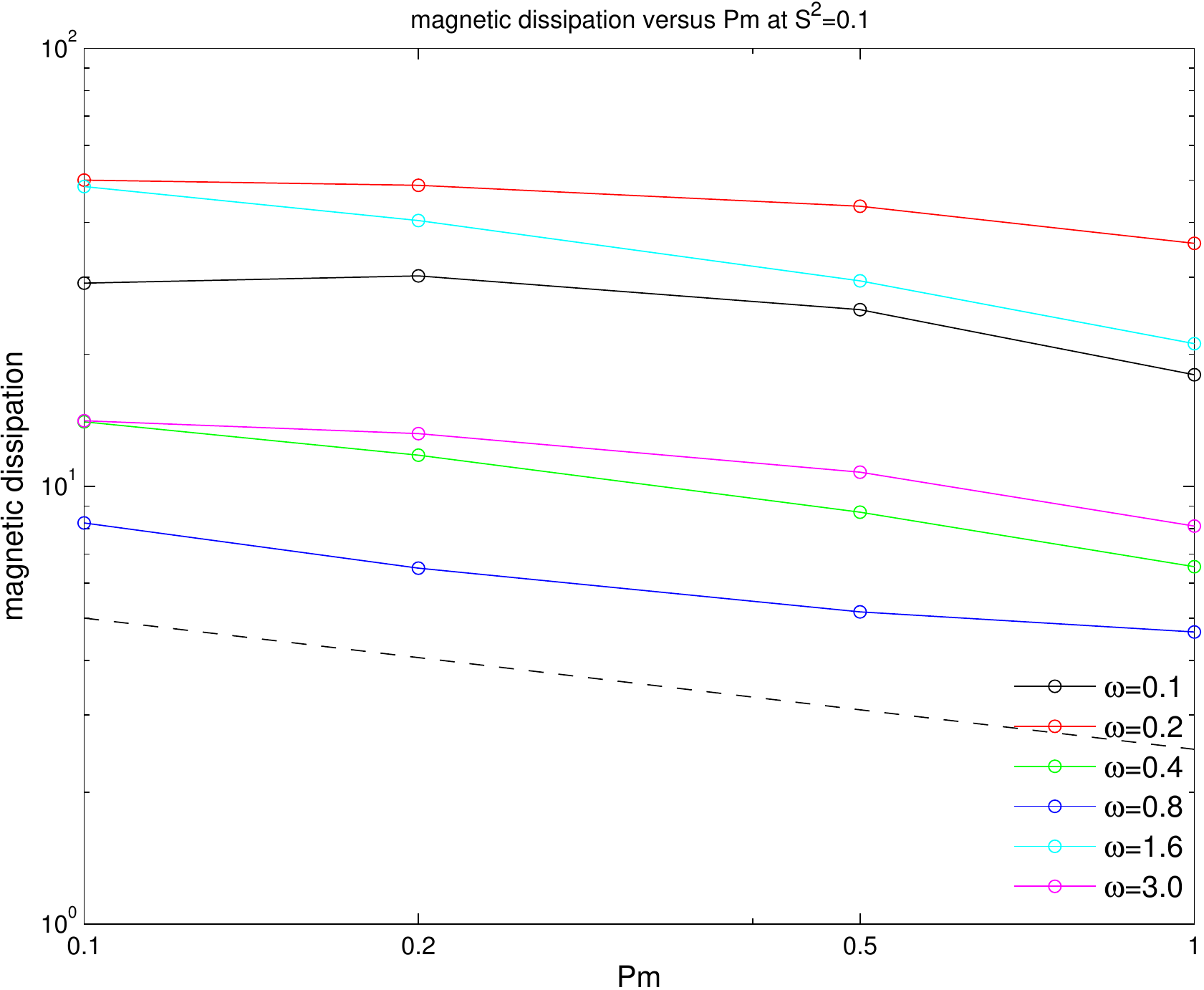}\label{fig7b}}
\caption{(a) viscous dissipation and (b) magnetic dissipation versus $Pm$ at various $\omega$. $S^2=0.1$. The two dashed lines denote the scaling laws $\propto Pm^{0.6}$ in (a) and $\propto Pm^{-0.3}$ in (b).}\label{fig7}
\end{figure}

To find the scaling laws about $Pm$, we plot the two dissipations versus $Pm$ as shown in Figure \ref{fig7}. The dependence on $Pm$ looks more regular than the dependence on $S$ in Figure \ref{fig5} because the field strength influences the resonant frequencies whereas diffusivities cannot. By taking the average of the slopes for different tidal frequencies we obtain the scaling laws
\begin{align}
\mbox{viscous dissipation}&\propto Pm^{0.6}, \nonumber\\
\mbox{magnetic dissipation}&\propto Pm^{-0.3}.
\end{align}
The dashed lines in Figures \ref{fig7a} and \ref{fig7b} denote respectively these two scaling laws. Then the ratio of magnetic to viscous dissipations is indeed close to $Pm^{-1}$ as shown in Figure \ref{fig6d}. It should be noted that our parameter regime is not the real regime, e.g. $E$ and $Pm$ are not too small, such that these scaling laws might not work in the real situation. With the lack of the knowledge about the physical properties in stars and planets, we do not know the real parameters, and even the estimation for the order of magnitude is not convincing. These scaling laws obtained in the moderate parameter regime at least give us some qualitative results that a moderate or stronger magnetic field improves both dissipations and the magnetic dissipation wins out viscous dissipation at a low $Pm$.

\subsection{A major result}

Now let us come back to the equations governing the system of magnetic tide. In equations \eqref{eq:ns}, \eqref{eq:bc} and \eqref{eq:b} there are four parameters, $E$, $S$, $Pm$ and $A$. In the linear regime we studied, the tidal dissipations should be proportional to $A^2$. Since we focus on the magnetic effect but not the rotational effect, Ekman number is not investigated but fixed to be a small value. Through the numerical studies we know the dependences on the two magnetic parameters $S$ and $Pm$, i.e. viscous dissipation is proportional to $S^2Pm^{0.6}$, magnetic dissipation to $S^2Pm^{-0.3}$, and total dissipation to $S^2$ but independent of $Pm$. Translating to the dimensional expression, we obtain the scaling law for the total dissipation about the strength of magnetic field,
\begin{equation}\label{eq:scaling}
\mbox{total dissipation}\propto B^2.
\end{equation}
Equation \eqref{eq:scaling} is a major result of this paper which has the astrophysical applications. For example, it can be used to {\bf estimate the strength of internal magnetic field} of the astronomical object of a binary system by observing the orbital evolution of the binary system. It should be noted that this scaling law is about magnetic field, and in addition to magnetic field, some other factors such as diffusivities, radius ratio, and so on can also influence the tidal dissipation. \citet{ogilvie2017} discussed the other factors in detail, and in the next section we will briefly discuss these other factors.

\section{Summary}

In this paper we numerically investigated the magnetic effect on the dynamical tide of a binary system. We tuned the tidal frequency, the field strength, and the two diffusivities to calculate both viscous and magnetic dissipations which are important for the orbital evolution of the binary system. It is found that a moderate or stronger magnetic field (in the sense that the Alfv\'en velocity is at least of the order of 0.1 of the surface rotational velocity) destroys the internal shear layers built by the propagation of inertial waves. Magnetic field modifies not only the flow structure but also the dispersion relation of waves excited by the tidal force such that the tidal resonance in rotating MHD is quite different from that in rotating flow, namely the resonance range is broadened by magnetic field to be out of $2\Omega$. Magnetic dissipation wins out viscous dissipation at the low tidal frequencies with a strong magnetic field but the two dissipations are comparable at the high tidal frequencies. The ratio of magnetic to viscous dissipations is almost inversely proportional to $Pm$ such that in the astrophysical situation at very low $Pm$ magnetic dissipation dominates over viscous dissipation with a moderate or stronger field, but the total dissipation does not depend on $Pm$. We obtained three scaling laws, viscous dissipation $\propto S^2Pm^{0.6}$, magnetic dissipation $\propto S^2Pm^{-0.3}$, and total dissipation $\propto S^2$. The last scaling law in its dimensional expression, total dissipation $\propto B^2$, can be applied to the estimation of the strength of internal magnetic field in the astronomical object of a binary system. 

Due to the limitation of our computational power, we do not compute the frequency-averaged dissipation which requires to scan a huge range of tidal frequency, especially in the case of a strong magnetic field. \citet{wei2016b} confirmed in a periodic box that the frequency-averaged dissipation is constant and interpreted this result with a simple damped harmonic oscillator. \citet{ogilvie2013} confirmed this result in a spherical shell for rotating flow and \citet{ogilvie2017} for rotating MHD.

To end this paper, we discuss some aspects which we did not consider in this work and can be studied in the future. Firstly, we focus on the two magnetic parameters $S$ and $Pm$ but did not investigate the other parameters, i.e. Ekman number $E$ measuring rotation and the radius ratio $R_i/R_o$ measuring the stellar or planetary structure. The dependence on $E$ is not very clear so far \citep{ogilvie-review}. The dependence of viscous dissipation in rotating flow on the core size was discussed in \citep{goodman2009} with the WKB analysis based on the critical latitude of inertial waves but this dependence is not clear in rotating MHD because magnetic field removes the critical latitude. These two parameters, $E$ and $R_i/R_o$, need to be further studied. Secondly, the imposed magnetic field in our model is uniform and vertical, which has the simplest geometry, and more complex field geometries should be further considered, e.g. dipolar and quadrupolar fields. In \citep{wei2010} various field geometries were studied in spherical Couette flow, namely flow between two differentially rotating spheres, and it was found that flow tends to be along field lines due to Alfv\'en's frozon-in theorem, and it can be inferred that the structure of tidal flow with various magnetic field geometries will be similar, namely flow along the field lines. Thirdly, our model is linear with the neglect of all the quadratic terms, i.e. $\bm u\cdot\bm\nabla\bm u$, $(\bm\nabla\times\bm b)\times\bm b$ and $\bm\nabla\times(\bm u\times\bm b)$. In \citep{wei2016a} it was found that the nonlinear inertial force suppresses the tidal response and dissipation near resonance. In \citep{ogilvie2014} it was also found that the dynamical tide in the nonlinear regime is quite different from that in the linear regime. So the nonlinear regime should also be further studied for comparison with the linear regime.

\section*{Acknowledgements}
Prof. Gordon Ogilvie and Dr. Yufeng Lin gave me valuable discussions during this work. An anonymous referee gave me good comments and suggestions and provided the information about the online published paper \citep{ogilvie2017}. The work was supported by the grant of 1000 youth talents of the Chinese government.

\bibliographystyle{apj}
\bibliography{paper}

\end{document}